\documentclass[preprint,12pt]{elsarticle}

\usepackage{amssymb}
\usepackage{amsmath}
\usepackage{multirow}

\usepackage{xurl}
\usepackage[linesnumbered,ruled,vlined]{algorithm2e}
\journal{Information and Software Technology}

\begin{document}

\begin{frontmatter}



\title{Enhancing User-Feedback Driven Requirements Prioritization}


\author[1]{Aurek Chattopadhyay} 
\author[2]{Nan Niu}
\author[3]{Hui Liu}
\author[4]{Jianzhang Zhang}

\affiliation[1]{organization={University of Cincinnati},
            city={Cincinnati},
            state={OH},
            country={USA}}
\affiliation[2]{organization={University of North Florida},
            city={Jacksonville},
            state={FL},
            country={USA}}
\affiliation[3]{organization={Beijing Institute of Technology},
city={Beijing},
country={China}
}
\affiliation[4]{organization={Hangzhou Normal University},
                city={Hangzhou},
                state={Zhejiang},
                country={China}}

\begin{abstract}

\noindent\textbf{Context:} Requirements prioritization is a challenging problem that is aimed to deliver the most suitable subset from a pool of candidate requirements. The problem is NP-hard when formulated as an optimization problem. Feedback from end users can offer valuable support for software evolution, and ReFeed represents a state-of-the-art in automatically inferring a requirement's priority via quantifiable properties of the feedback messages associated with a candidate requirement.

\noindent\textbf{Objectives:} In this paper, we enhance ReFeed by shifting the focus of prioritization from treating requirements as independent entities toward interconnecting them. Additionally, we explore if interconnecting requirements provide additional value for search-based solutions. 

\noindent\textbf{Methods:} We leverage user feedback from mobile app stores to group requirements into topically coherent clusters. Such interconnectedness, in turn, helps to auto-generate additional ``requires'' relations in candidate requirements. These ``requires'' pairs are then integrated into a search-based software engineering solution. 

\noindent\textbf{Results:} The experiments on 94 requirements prioritization instances from four real-world software applications show that our enhancement outperforms ReFeed. In addition, we illustrate how incorporating interconnectedness among requirements improves search-based solutions.

\noindent\textbf{Conclusion:} Our findings show that requirements interconnectedness improves user feedback driven requirements prioritization, helps uncover additional ``requires'' relations in candidate requirements, and also strengthens search-based release planning. 

\end{abstract}




\end{frontmatter}

\section{Introduction}

Deciding which requirements really matter is a challenging task. Tight time-to-market constraints and practical budget limitations demand effective and efficient requirements prioritization solutions. Modern software industry invests heavily in strategic release planning~\cite{Marner-Computers22, Spinellis-SW15, Nayebi15, Zorn-Pauli-REFSQ13}, where a special case is known as the \emph{next release problem} (NRP). Given a set of candidate requirements, the NRP is concerned with prioritizing a subset of requirements over the others, so that the prioritized part gets implemented and delivered, whereas the remaining requirements would not be included in the upcoming version of the software.

The literature has offered much search-based software engineering (SBSE) support for the NRP~\cite{Bagnall-IST01, Feather-RE02, Zhang-GECCO07, Finkelstein-REJ09, Durillo-ESE11, Cai12, Xuan-TSE12, Karim-SSBSE14}. Assuming there are $n$ candidate requirements, the NRP solutions are bound in a space of 2$^{n}$. Thus, metaheuristic search guided by some objective function could identify high quality but possibly suboptimal solutions in a computationally tractable way. The objective function is defined by commonly referring to some attributes of the requirements. For example, Bagnall \emph{et al.}~\cite{Bagnall-IST01} aimed to maximize the value of different stakeholders' desired requirements while ensuring that the implementation cost would not exceed a certain limit.

\begin{figure*}[!t]
\centering
\includegraphics[width=\linewidth]{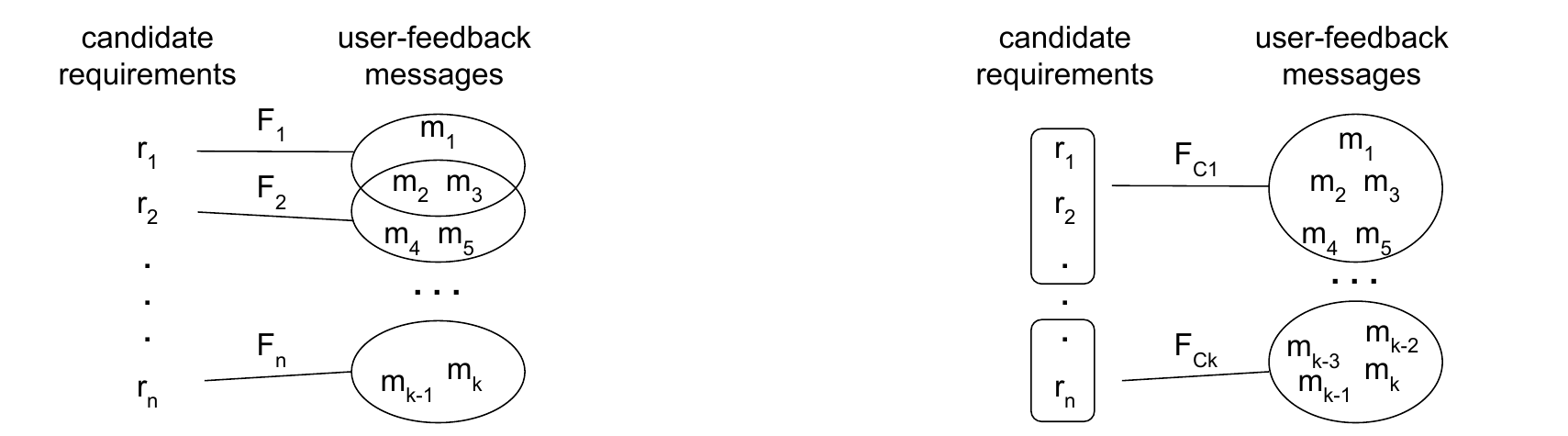}
\flushleft{\mbox{\textbf{\footnotesize{$\qquad\quad\;\;\;$(a) $\;$ReFeed$\qquad\qquad\qquad\qquad\qquad\qquad\qquad\,\,$(b) $\;$iReFeed}}}}
\caption{Compared with ReFeed~\cite{Kifetew-IST21} that associates user-feedback messages to a single requirement, our approach of iReFeed associates messages to a cluster of \emph{interconnected} requirements. The ``i'' in iReFeed emphasizes this interconnectedness.}
\label{fig:1}
\end{figure*}

An important attribute relates to the feedback from the end users of the software application. This data source is valuable for understanding what users request, how they like or dislike the software, etc. In a seminal paper, Kifetew \emph{et al.}~\cite{Kifetew-IST21} introduced ReFeed, a user-feedback driven requirements prioritization method. ReFeed associates user-feedback to requirements, and then computes the requirements priorities based on the extracted properties of the associated feedback. Fig.~\ref{fig:1}-a illustrates ReFeed which we review in more detail later.

Inspired by ReFeed, we propose a novel approach to enhance user-feedback driven requirements prioritization. Fig.~\ref{fig:1}-b highlights the key change. Rather than treating requirements as independent entities, we leverage user-feedback messages to group the requirements into topically coherent clusters. In this way,  the messages' influences on the priorities are no longer done at each requirement's level, but at a granularity of interconnected requirements. We call our approach iReFeed to reflect the interconnectedness of the requirements in the same topic cluster. In Fig.~\ref{fig:1} the set of candidate requirements is denoted by \{$r_1,r_2,....,r_n$\}, where n is the number of candidate requirements. The set of user-feedback messages is denoted by \{$m_1,m_2,....,m_p$\}, where p is the size of user-feedback messages. In ReFeed, $F_i$ represents the set of feedback messages associated with the candidate requirement $r_i$. In iReFeed, $F_{C_j}$ represents the set of feedback messages associated with the j-th candidate requirement cluster $C_j$, where j= 1,....k.

Recognizing the requirements' interconnectedness is important for prioritization. Karim and Ruhe~\cite{Karim-SSBSE14}, for instance, introduced ``theme'' where the requirements within a theme would be preferably delivered or postponed together. Aydemir \emph{et al.}~\cite{Aydemir-RE18} further extended it to goal modeling, e.g., by including all the leaf goals of a theme in the next release or excluding them altogether. While theme models a ``requires'' relation symmetrically (i.e., ``req$_1$ requires req$_2$, \emph{and} req$_2$ requires req$_1$'')~\cite{Carlshamre-RE01}, the asymmetric ``requires'' is more common and practically influential to prioritization. Such one-way ``requires'' manifests itself in different ways, including functional dependency~\cite{Carlshamre-RE01}, temporal relation~\cite{Robinson-CSUR03}, and refinement structure~\cite{Aydemir-RE18}.

To explore the extent to which iReFeed is conducive to ``requires'' identifications, we exploit large language models (LLMs)---OpenAI's ChatGPT 4.5 and 4o in particular---to investigate whether the topic clusters resulted from our approach would lead to the generation of more ``requires'' pairs. Surprisingly, compared to feeding ChatGPT with all the requirements once, focusing on the requirements within iReFeed's topic cluster uncovered additional ``requires'' pairs. These pairs, in turn, helped improve a state-of-the-art SBSE solution to the NRP.

This paper makes three main contributions.

\begin{itemize}
\item We improve ReFeed by incorporating user-feedback from mobile app stores in interconnecting requirements, and the experiments on four real-world applications show that our approach consistently achieves better prioritization results than ReFeed.
\item To examine the usefulness of our approach's requirements interconnectedness, we prompt ChatGPT on the Word Processor NRP benchmark data~\cite{Karim-SSBSE14} and reveal that additional ``requires'' pairs could be generated with the help of our topic-cluster results.
\item We demonstrate the added value of the ``requires'' pairs through a well-suited SBSE NRP solution, namely NSGA-II, on the Word Processor benchmark.
\end{itemize}

We share our study materials and results publicly at \url{https://doi.org/10.5281/zenodo.18881986} to facilitate replication. The rest of the paper is organized as follows. Section~\ref{sect:2} describes the context of related work in which the current paper is located. Section~\ref{sect:3} presents our iReFeed approach. Section~\ref{sect:4} evaluates iReFeed against ReFeed. Section~\ref{sect:5} reports our ChatGPT experiments on identifying ``requires'' pairs. Section~\ref{sect:6} illustrates the incorporation of the ``requires'' pairs into the NSGA-II algorithm. Section~\ref{sect:7a} discusses the threats to validity. Section~\ref{sect:7} draws the conclusion and points out future work directions.

\section{Background and Related Work}\label{sect:2}

\subsection{SBSE for Automating Requirements Prioritization}\label{subsect:21}

Deciding which, among a set of requirements, are to be considered first is a strategic process in software development~\cite{Perini-TSE13, ruhe10}. Traditionally, requirements prioritization is done manually, and hence is effective for only a small number of requirements. The manual approaches include win-win negotiation for deriving a final requirements ranking from an agreement among subjective evaluations by different stakeholders~\cite{Ruhe-SEKE02}, the 100-point method that lets stakeholders distribute their points individually before aggregating the points for ranking the requirements~\cite{leffingwell03}, and the analytic hierarchy process (AHP) in which the decision maker compares every pair of requirements in terms of value and cost, feeding into the eigenvalue calculations in order to determine the requirements priorities~\cite{Karlsson-SW97}. Karlsson \emph{et al.}~\cite{Karlsson-IST98} compared several traditional methods, and showed that AHP was the most promising though limited in scalability. For $n$ requirements, AHP requires $\frac{n \times (n-1)}{2}$ comparisons in one dimension (e.g., value). Perini \emph{et al.}~\cite{Perini-REW07} reported that AHP suffers scalability issues with requirements sets larger than about 50.

\begin{table*}[!t]
\centering
\caption{Snippet of the Word Processor Data (\emph{adopted from Karim and Ruhe's work~\cite{Karim-SSBSE14} with the data linked in~\cite{WP-Data-Web})}}\label{table:WP}
\small
\resizebox{\textwidth}{!}{     
\begin{tabular}{| l | c | c | c | c | c | c | c | c |}
\hline
\multirow{2}{*}{\bf $\qquad\,$Requirement} & \multicolumn{4}{c|}{\bf Value Scores} & \multicolumn{4}{c|}{\bf Resource Estimates} \\
\cline{2-9}
& {\bf Stakeholder1} & {\bf Stakeholder2} & {\bf \ldots} & {\bf Stakeholder4} & {\bf \ldots} & {\bf Design} & {\bf Development} & {\bf QA / Testing} \\
\hline
$r_1$: Create a new file & 8 & 9 & \ldots & 9 & \ldots & 17 & 22 & 12 \\
$r_2$: Open an existing file & 8 & 9 & \ldots & 9 & \ldots & 20 & 25 & 13 \\
$r_3$: Close current file & 8 & 9 & \ldots & 1 & \ldots & 5 & 1 & 2 \\
\multirow{2}{*}{\ldots $\qquad$ \ldots $\qquad$ \ldots} & \multirow{2}{*}{\ldots} & \multirow{2}{*}{\ldots} & \multirow{2}{*}{\ldots} & \multirow{2}{*}{\ldots} & \multirow{2}{*}{\ldots} & \multirow{2}{*}{\ldots} & \multirow{2}{*}{\ldots} & \multirow{2}{*}{\ldots} \\
 & & & & & & & & \\
$r_{49}$: Load help file & 2 & 7 & \ldots & 6 & \ldots & 3 & 3 & 3 \\
$r_{50}$: Search a text in the & \multirow{2}{*}{1} & \multirow{2}{*}{7} & \multirow{2}{*}{\ldots} & \multirow{2}{*}{6} & \multirow{2}{*}{\ldots} & \multirow{2}{*}{2} & \multirow{2}{*}{2} & \multirow{2}{*}{2} \\
$\quad\;\;$ help file & & & & & & & & \\
\hline
\end{tabular}}
\normalsize
\end{table*}

In pursuit of scalable solutions, researchers have explored computational search, especially metaheuristic search. Bagnall \emph{et al.}~\cite{Bagnall-IST01} were among the first to connect requirements prioritization with SBSE by defining the NRP, i.e., the problem of selecting an optimal set of requirements to be delivered in the software system's next stable version. Here, optimality means satisfying the stakeholder demands as much as possible, and meanwhile ensuring that there are enough resources to undertake the necessary development. This optimization problem is shown to be NP-hard~\cite{Bagnall-IST01}. To illustrate Bagnall \emph{et al.}'s NRP formulation, we show in Table~\ref{table:WP} a snippet of the Word Processor data~\cite{Karim-SSBSE14}.

This benchmark contains 50 requirements, some of which are listed in the left column of Table~\ref{table:WP}. The solution to the NRP is given by the decision vector: $\vec{x}$=$\{x_1, x_2, \ldots, x_{50}\}$, $x_i \in \{0, 1\}$ for 1$\le$$i$$\le$50. In this vector, $x_i$=1 if requirement $r_i$ is selected to be part of the next release, and $x_i$=0 otherwise. To determine if a requirement should be selected or not, one shall consider some critical attributes and their values. In Table~\ref{table:WP}, columns 2--5 illustrate the stakeholder value scores, and columns 6--9 display the implementation-related resource estimates. Bagnall \emph{et al.}~\cite{Bagnall-IST01} defined the search objective to: Maximize $\sum_{i=1}^{50} x_i \cdot \mathrm{value}(r_i)$, subject to $\sum_{i=1}^{50} x_i \cdot \mathrm{cost}(r_i) \le B$, where $\mathrm{value}(r_i)$ is the (weighted) sum of stakeholders' values placed on $r_i$, $\mathrm{cost}(r_i)$ is the total resource estimates of implementing $r_i$, and $B$ is some bound. Thus, SBSE techniques like hill climbing and simulated annealing could be used to find a high quality, but possibly suboptimal, solution $\vec{x}$~\cite{Bagnall-IST01}.

Although Bagnall \emph{et al.}~\cite{Bagnall-IST01} originally presented a single objective formulation of the NRP, others have adopted multi-objective optimization approaches~\cite{Finkelstein-REJ09,Cai12,Karim-SSBSE14,Zhang-GECCO07}. Karim and Ruhe~\cite{Karim-SSBSE14} showed that maximizing individual requirement's value and theme-based coherence are conflicting objectives. Similarly, Finkelstein \emph{et al.}~\cite{Finkelstein-REJ09} revealed the trade-offs between maximizing the total value and being fair (e.g., reducing the variance of the number of implemented requirements for each stakeholder).

In a multi-objective optimization problem, we are often interested in finding Pareto-optimal solutions. Suppose there are $M$ objective functions. A decision vector $\vec{x}$ is Pareto-optimal if there is no $\vec{y}$ that is better than $\vec{x}$ in at least one $i$=1, 2, \ldots, $M$, and no worse than $\vec{x}$ in the others. Let us consider two simple solutions about the Word Processor: $\alpha$ selects only $r_1$ to release, and $\beta$ selects only $r_2$. With the data given in Table~\ref{table:WP}, $\alpha$ is better than $\beta$ in resource consumption (i.e., it costs less to implement $r_1$ than $r_2$), and meanwhile, $\alpha$ is no worse than $\beta$ in stakeholder values. Therefore, in light of $\alpha$, $\beta$ is not a Pareto-optimal solution. Among the many search algorithms investigated in the NRP literature, NSGA-II is well-suited~\cite{Karim-SSBSE14,Cai12,Zhang-GECCO07}, exhibits the best run time~\cite{Rahimi-ASC23,Zhang-TOSEM18}, and performs the best in finding Pareto-optimal solutions~\cite{Finkelstein-REJ09,Durillo-ESE11,Zhang-TOSEM18}.

In summary, SBSE offers strategic and automatic ways to solve the NRP. Extensive empirical studies show that one of the best search-based NRP algorithms is NSGA-II, which we integrate with the knowledge obtained from our iReFeed approach.

\subsection{Interrelated Requirements in Prioritization}\label{subsect:22}

In addition to stakeholder values and effort estimates, the requirements interrelations are among the most influential factors of release planning models used in industry~\cite{Svahnberg-IST10, Ameller-PROFES16}. Carlshamre \emph{et al.}~\cite{Carlshamre-RE01} observed several relation types, such as ``AND'' and ``requires''. The distinction is that ``AND'' suggests a requires-relation in both ways whereas ``requires'' indicates the relation in only one way. For example, a printer requires a driver to function, and the driver also requires the printer to function~\cite{Carlshamre-RE01}. This is an instance of ``AND'' relation.

In contrast, emailing a scanned document requires a network connection, but not the other way around~\cite{Carlshamre-RE01}. Here, a ``requires'' relation exists between the pair of requirements. Notably, ``requires'' embodies other forms than functional dependency, e.g., ``auto-logout requires a detected inactivity period, but not the opposite'' is a temporal relation~\cite{Robinson-CSUR03}, and ``freight management improvement'' is refined by ``freight order history logging''~\cite{Aydemir-RE18}. In all these cases, we use $r_a \to r_b$ to denote ``$r_a$ requires $r_b$''. The implication is that $r_b$ shall not be prioritized lower or later than $r_a$. In case of ``AND'', i.e., $r_a \to r_b$ \emph{and} $r_b \to r_a$, both requirements share the same priority.

Referring to the Word Processor requirements of Table~\ref{table:WP}, we have $r_3 \to r_2$. Now consider three decisions: $\beta = \{r_2\}$, $\gamma = \{r_3\}$, and $\delta = \{r_2, r_3\}$, which would respectively release $r_2$ only, $r_3$ only, and both together. Here, $\gamma$ is not congruent with $r_3 \to r_2$, and hence should not be regarded as a valid NRP solution. One can impose the ``requires'' constraints in a post-processing step of SBSE, or encode them as part of the SMT/OMT solving~\cite{Aydemir-RE18}.

In short, ``requires'' represents a common and critical relation that influences the quality of prioritization outcomes. However, such relations were identified manually in prior studies~\cite{Carlshamre-RE01, Aydemir-RE18}. We exploit ChatGPT for automating the ``requires'' identification, and further investigate our approach's impact on uncovering the ``$r_a \to r_b$'' pairs. 

\subsection{CrowdRE and ReFeed}\label{subsect:23}

Not only are requirements themselves interrelated, but they also relate to other data and specifically to the feedback from the end users of the software. Groen \emph{et al.}~\cite{Groen-REFSQ15} defined crowd-based requirements engineering (RE), or CrowdRE, as a semi-automated approach for obtaining and analyzing any kind of user feedback from a pool of current and potential stakeholders. Issue reports, forum discussion, and app reviews are among the common data sources for CrowdRE.

An earlier tool, ChangeAdvisor~\cite{Palomba-ICSE17}, experimented three topic modeling algorithms (namely LDA, LDA-GA, and HDP) for clustering user feedback into groups expressing similar needs. ChangeAdvisor adopted HDP as it provided a good trade-off between quality and execution time. A subsequent tool, CLAP~\cite{Scalabrino-TSE19}, trained a random forest classifier to differentiate user reviews into categories like feature and bug. CLAP further involved clustering, e.g., by using DBScan to group the user reviews reporting the same bug in one cluster. CLAP finally employed another random forest to classify whether a cluster would be high priority or low priority. Recently, {\sc Radiation}~\cite{Nayebi-RE23} was introduced to recommend the removal of certain user interface functionality. {\sc Radiation} applied HDP to cluster user reviews, followed by the use of a random forest classifier to determine what to delete based on quantifiable properties such as review numbers, ratings, and sentiments.

Although other CrowdRE methods exist, the above tools illustrate that topic modeling and clustering are effective in handling the high volume of feedback data and that the review aggregates are valuable for software evolution~\cite{Stronstad-REW23}. Nevertheless, ChangeAdvisor traces user feedback to source code, CLAP prioritizes user reviews instead of candidate requirements, and {\sc Radiation} recommends to remove existing user interface features but falls short of potentially enriching the software with better features.

Kifetew \emph{et al.}~\cite{Kifetew-IST21}  pioneered the use of feedback in automatically prioritizing requirements, which inspires our research. As outlined in Fig.~\ref{fig:1}-a, ReFeed comprises three major steps: associating feedback with each requirement, extracting feedback properties, and inferring the priority of requirements. The feedback-requirement associations are established on the basis of textual similarity. ReFeed computes the Jaccard similarity between each feedback message and every requirement, and then forms associations if the similarity is greater than a threshold. Similarity-based associations are common in contemporary CrowdRE tools~\cite{Palomba-ICSE17, Nayebi-RE23}.

ReFeed extracts quantifiable properties from user feedback that are relevant for prioritizing the requirements. These properties result mainly from sentiment analysis and speech-acts analysis~\cite{Kifetew-IST21}. Let $N$ be the total number of sentences in a feedback message, and $N_{pos}$ (or $N_{neg}$) be the number of sentences with positive (or negative) sentiment. Due to neutral sentiment, $N_{pos} + N_{neg}$ does not always equal to $N$. Given a user-feedback message, $neg$ = $\frac{\sum_{n=1}^N SentimentScore_{neg}(Sentences[n])}{N_{neg}}$ and $pos$ = $\frac{\sum_{n=1}^N SentimentScore_{pos}(Sentences[n])}{N_{pos}}$ quantify the message's negative and positive sentiments respectively. ReFeed uses StanfordCoreNLP to compute $SentimentScore$ at the sentence level~\cite{Kifetew-IST21}. The speech-acts analysis is aimed to calculate the extent a feedback message conveys feature request as opposed to other intentions like bug fix. To that end, a message's intention score, $int$, is calculated as the average of the intentions of its sentences. An $int$=1 implies the message is a feature request, whereas a lower $int$ score suggests that the feedback is less about making a feature request. In this formulation, feature request intention receives a higher intention score and contributes more strongly to the prioritization as opposed to other intentions like bug fix.

Having associated each requirement $r$ via $F$ with a set of feedback messages, Kifetew \emph{et al.}~\cite{Kifetew-IST21} inferred the requirement's priority $P$ in ReFeed as\footnote{We ignore the severity measure from~\cite{Kifetew-IST21} as it is based purely on the negative sentiment of feedback. As can be seen in equation (\ref{eq:1}), the negative sentiment has already been considered in ReFeed.}:

\begin{equation} \label{eq:1}
P_r = \frac{\sum_{i=1}^{|F|}{[sim(r, F[i]) * (neg_{F[i]} + pos_{F[i]} + int_{F[i]})]}}{|F|} 
\end{equation}

\noindent where $F[i]$ denotes the feedback message at position $i$ in the set mapped by $F$, and $sim(r, F[i])$ represents the textual similarity value between the requirement $r$ and the message $F[i]$. If $|F|$=0, then no feedback is associated with $r$ and that requirement's priority is set to be 0. Intuitively, equation (\ref{eq:1}) assigns a higher priority to $r$ if its feedback is more similar to the requirement, embodies stronger sentiment, and has a greater intention to convey feature request.

In essence, ReFeed offers a fully automatic way to prioritize requirements based on user feedback. Improving ReFeed with requirements interconnectedness is precisely the focus of our research.

\section{A New Approach: iReFeed}\label{sect:3}

\begin{figure}[!b]
\vspace*{-0.25em}
\centering
\includegraphics[width=1.00\columnwidth]{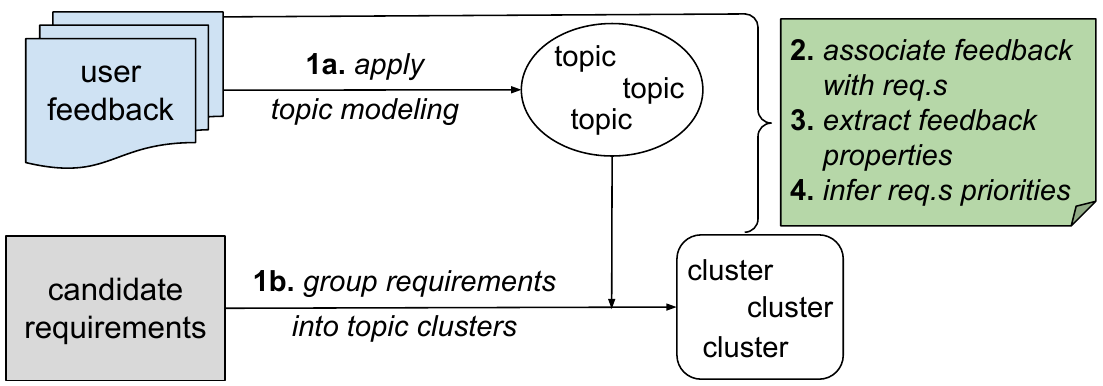}\\
\vspace*{-0.50em}
\caption{Overview of iReFeed processing pipeline.}
\label{fig:2}
\end{figure}

We present in Fig.~\ref{fig:2} an overview of our iReFeed approach, while Algorithm~\ref{alg:irefeed} summarizes the steps. The figure elaborates our newly proposed steps of performing topic modeling on user feedback and then grouping the requirements into topically coherent clusters. Even though iReFeed shares with ReFeed the overall steps of associating feedback with requirements, extracting feedback properties, and inferring requirements priorities, important differences exist. We discuss these differences, together with key design rationales and implementation details, in this section.

\vspace*{+1.0em}

\noindent {\bf Step 1:} Our construction of clusters of interconnected requirements consists of topic modeling of user feedback and then grouping the candidate requirements by using the resulting topics. The primary reason for applying topic modeling to feedback is due to the quantity of the data. Typically, there are tens to hundreds of candidate requirements, which can be inadequate for insightful topics to emerge in the latent space. In contrast, tens or hundreds of thousands of feedback messages can be readily collected for a software application, creating a sizeable critical mass for applying topic modeling.

We implement two topic modeling methods in iReFeed: LDA and BERTopic, due to their viable applications in CrowdRE~\cite{Silva-ESE21, Sihag-RE23}. A salient distinction is that LDA uses the corpus from the tokens of the user feedback data, while BERTopic exploits a pre-trained transformer model with data crawled from external sources like Wikipedia. The literature has not recommended any common number of topics to be selected~\cite{Silva-ESE21}. Thus, we experimented with several numbers of topics, and selected 20 as an appropriate one to use in our current work. Previous studies~\cite{Hindle2015, Tiarks2014} have also found 20 topics to be appropriate. We trained LDA on user reviews with 15 passes to ensure convergence, and then converted each candidate requirement into a bag-of-words representation to cluster them based on the highest topic distributions. For BERTopic, we applied UMAP (number of components=5, number of neighbors=15, minimum distance=0) for dimensionality reduction and HDBSCAN (minimum number of samples=10) for clustering user feedback into topics. We then converted the requirements into sentence-BERT embeddings and assigned them to the most relevant topic cluster based on the embeddings' cosine similarities. Note that the feedback topics are important intermediate results, because they serve as a means to form clusters of interconnected requirements.

\vspace*{+0.75em}
\noindent {\bf Step 2:} As in~\cite{Kifetew-IST21, Nayebi-RE23}, we exploit textual similarity to associate feedback with requirements. In~\cite{Palomba-ICSME15}, for instance, Palomba \emph{et al.} used the cosine similarity threshold of 0.6 for the purpose of associations; yet, Nayebi \emph{et al.}~\cite{Nayebi-RE23} adjusted the threshold to 0.65 in order to achieve a more accurate matching level. Similarly, we slightly increased the Jaccard similarity threshold of 0~\cite{Kifetew-IST21} to the cosine similarity of 0.1 to obtain associations in iReFeed and in our ReFeed implementation.

\begin{algorithm}[H]
\label{alg:irefeed}
\SetKwInput{KwInput}{Input}                
\SetKwInput{KwOutput}{Output}              
\DontPrintSemicolon
  
  \KwInput{CandidateRequirements: R, UserFeedback: UF}
  \KwOutput{Candidate requirements ranked by priority}
 
  \SetKwFunction{FMain}{Main}
  \SetKwFunction{FExtractFeedbackProperties}{ExtractFeedbackProperties}
 
  \SetKwProg{Fn}{Function}{:}{}
  \Fn{\FExtractFeedbackProperties{$Feedback$}}{

        \For{each feedback message f in Feedback}{
            	 posSent[f] $\leftarrow$ ComputePositiveSentiment(f)\; 
                negSent[f] $\leftarrow$ ComputeNegativeSentiment(f)\;
	            int[f] $\leftarrow$ ComputeIntention(f)\;
        }

        \KwRet FeedbackProperties\;
  }
  \;

  \SetKwProg{Fn}{Function}{:}{\KwRet}
  \Fn{\FMain}{
        topics $\leftarrow$ FeedbackTopicModeling(UF)\;
        RequirementClusters $\leftarrow$ GenerateTopicClusters(R, topics)\;
        \For{each cluster c in RequirementClusters}{
            AssociatedFeedback $\leftarrow$ FindAssociatedFeedback(c, UF)\;
            FeedbackProperties $\leftarrow$ ExtractFeedbackProperties(AssociatedFeedback)\;
            \For{each requirement r in c}{ 
		    
		      priority[r] $\leftarrow$ ComputePriorityScore(r, FeedbackProperties)\;

        }}

        \KwRet SortDescending(R, priority)\;
  }
\caption{iReFeed}
\end{algorithm}
\vspace*{+0.75em}       

Different from ReFeed, the associations are established at a requirements cluster level in our work. Consider an example of two requirements and three feedback messages. If their cosine similarities are: $sim(r_1, m_1)$=0.15, $sim(r_1, m_2)$=0.2, $sim(r_1, m_3)$=0.06, $sim(r_2, m_1)$=0.07, $sim(r_2, m_2)$=0.22, and $sim(r_2, m_3)$=0.13, then under a 0.1 threshold, ReFeed would associate \{$r_1$\} with \{$m_1$, $m_2$\} and associate \{$r_2$\} with \{$m_2$, $m_3$\}. In iReFeed, if $r_1$ and $r_2$ are in the same cluster, then we link \{$r_1$, $r_2$\} with \{$m_1$, $m_2$, $m_3$\} by taking the union of the messages associated with the cluster's constituent requirements. In this way, for instance, the influence of $m_1$ on prioritizing $r_2$ is incorporated in iReFeed but neglected in ReFeed. The influence, as will be seen later, does get moderated by the similarity score.

\vspace*{+1.0em}
\noindent {\bf Step 3:} To extract quantifiable properties of the feedback messages, we perform sentiment analysis via the Stanza library from the Stanford NLP package~\cite{Qi-arXiv20}. Similar to ReFeed, we compute negative sentiment and positive sentiment separately. These computations are done first at a sentence level, and then aggregated to a message level.

Our calculation of an intention score adopts a review classification method, rather than the user feedback ontology used in~\cite{Kifetew-IST21}. While the ontology may need to be updated from one domain to another, a classifier trained on diverse review data can offer an automatic alternative to recognize the type of feedback. To that end, we built a random forest classifier on top of the datasets shared by Scalabrino et al.~\cite{Scalabrino-TSE19}. The datasets contain 725 reviews from 14 apps, and the reviews are labeled with six classes: feature, bug, performance, usability, security, and energy. Most of the reviews belong to the categories of bug or feature. We developed the random forest with a train-test split ratio of 80-20, and achieved an accuracy of 79\%. This performance is comparable to the random forest classifiers used in the CrowdRE literature~\cite{Scalabrino-TSE19, Nayebi-RE23}. Note that we applied the same sentiment analysis and content classification to ReFeed and iReFeed, so there is no difference in extracting feedback properties.

\vspace*{+1.0em}
\noindent {\bf Step 4:} The chief difference in inferring requirements priority is illustrated in Fig.~\ref{fig:1}-b. In iReFeed, the feedback mapping, $F_C$, is no longer from a single requirement, but from a cluster. Therefore, a priority of a requirement $r$ in iReFeed is defined as:

\small
\begin{equation} \label{eq:2}
P_r = \frac{\sum_{i=1}^{|F_C|}{[sim(r, F_C[i]) * (neg_{F_C[i]} + pos_{F_C[i]} + int_{F_C[i]})]}}{|F_C|} 
\end{equation}
\normalsize

\noindent where $F_C$ represents the feedback mapping from the cluster that $r$ is in, and the rest of the notations have the same interpretations in equation (\ref{eq:1}). Orthogonal to LDA and BERTopic, we design a mechanism to weigh in the coherence of iReFeed's requirements clusters. We call it LDA-C or BERTopic-C, and define a requirement's priority in the C variants as:

\footnotesize
\begin{equation} \label{eq:3}
P_r = \frac{\sum_{i=1}^{|F_C|}{[\alpha(C_j) * sim(r, F_C[i]) * (neg_{F_C[i]} + pos_{F_C[i]} + int_{F_C[i]})]}}{|F_C|} 
\end{equation}
\normalsize

\noindent where $C_j$ represents the requirement cluster containing requirement r, and the newly added term $\alpha(C_j) = $(1 + average pairwise similarity of $\forall r_i, r_j \in C_j$). This factor boosts the priorities of the requirements in an internally coherent cluster; however, due to the constant 1, $\alpha(C_j)$ does not penalize the poorly coherent clusters. In fact, when $\alpha(C_j)$=1, equation (\ref{eq:3}) is reduced to equation (\ref{eq:2}), implying that requirements cluster's coherence is expected to influence the prioritization results in a more general sense.

\section{Evaluating Prioritization Results}\label{sect:4}

The research question that we investigate in this section is 
{\bf RQ$_1$: ``How does iReFeed compare to ReFeed in delivering requirements prioritization results?''} We treat ReFeed as a state-of-the-art method of using feedback to prioritize candidate requirements, and compare ReFeed with the four variants of iReFeed: LDA, BERTopic, LDA-C, and BERTopic-C.

\subsection{Datasets and Metrics}\label{subsect:41}

\begin{table*}[!t]
\centering
\caption{Characteristics of RQ$_1$ Datasets}\label{table:Char}
\small
\resizebox{\textwidth}{!}{
\begin{tabular}{ | c c | c c | c c |}
\hline
\multicolumn{1}{|c}{\bf app [source of} & \multicolumn{1}{c|}{\bf \# of prioriti-} & \multicolumn{1}{c}{\bf \# of re-} & \multicolumn{1}{c|}{\bf requirements} & \multicolumn{1}{c}{\bf \# of user-} & {\bf feedback} \\
\multicolumn{1}{|c}{\bf release notes]} & \multicolumn{1}{c|}{\bf zation instances} & \multicolumn{1}{c}{\bf quirements} & \multicolumn{1}{c|}{\bf time period} & \multicolumn{1}{c}{\bf feedback messages} & {\bf time period} \\
\hline \hline
Discord~\cite{DiscordReleaseNotes-Web} & 18 & 373 & July 2020--March 2025 & 368,367 & Jan 2020--March 2025 \\
Microsoft 365 Word~\cite{Microsoft365ReleaseNotes-Web} & 12 & 295 & July 2018--Dec 2024 & 143,547 & Jan 2018--Dec 2024\\
Webex~\cite{WebexReleaseNotes-Web} & 29 & 360 & Feb 2022--July 2024 & $\;$17,829 & Jan 2022--July 2024 \\
Zoom~\cite{ZoomReleaseNotes-Web} & 35 & 951 & Feb 2022--March 2025 & $\;$62,074 & Jan 2022--March 2025 \\
\hline
\end{tabular}}
\normalsize
\end{table*}

To answer RQ$_1$, we constructed 94 requirements prioritization instances from four real-world software applications: Discord, Microsoft 365 Word, Webex, and Zoom. Table~\ref{table:Char} lists some characteristics of our evaluation datasets. We relied on the release notes maintained by the software vendors themselves to identify requirements. In addition, we collected the release time of each requirement. Finally we combined the requirements from two consecutive release periods to form a prioritization instance.

\begin{table*}[!t]
\centering
\caption{Excerpts from One Prioritization Instance of the Zoom Dataset (the top 19 requirements were released in February 2025, and the bottom 52 requirements were released in March 2025)}\label{table:OneInstance}
\normalsize        
\resizebox{\textwidth}{!}{      
\begin{tabular}{ | r | l | c |}
\hline
\multicolumn{1}{|c}{\bf \#} & \multicolumn{1}{|c|}{\bf Feature Description} & $\!\!${\bf In Ground Truth?} \\
\hline \hline
1 & Meeting participants can request access to recordings directly from the meeting card or recording  & Yes \\
& $\qquad$ link without having to send separate messages through chat, email, or calls. Meeting hosts &\\
& $\qquad$ receive these requests through in-product notifications and emails, which direct them to a &\\
& $\qquad$ viewing page where they can manage access permissions. Hosts can view all requesters on the &\\
& $\qquad$ share modal, grant or deny access, and manage existing permissions for specific users. The &\\
& $\qquad$ feature respects all security settings, including authentication requirements and password protection. & \\
2 & After a meeting ends, a dynamic pop-up appears, based on the assets utilized during the meeting,  & Yes \\
& $\qquad$ directing users to the meeting details page. Here, they can access available meeting assets  & \\
& $\qquad$ such as meeting recording, summary, continuous chat log, as well as any content shared during  & \\
& $\qquad$ the session, such as whiteboards, links, and notes. & \\
& \ldots $\qquad\qquad$ \ldots $\qquad\qquad$ \ldots $\qquad\qquad$ \ldots $\qquad\qquad$ \ldots $\qquad\qquad$ \ldots $\qquad\qquad$ \ldots $\qquad\qquad$ \ldots & \ldots \\
19 & Users can view a complete list of meeting participants directly within the primary meeting card interface.  & Yes \\
& $\qquad$ The participant list appears in the full meeting page, providing clear visibility of all attendees. & \\
\hline
1 & Hosts and co-hosts can move participants directly from the waiting room to designated breakout & No \\
& $\qquad$  rooms without requiring them to enter the main meeting room first. Additionally, they can& \\ 
& $\qquad$  lock breakout rooms to prevent participants from returning to the main session. This feature & \\
& $\qquad$  works for both signed-in Zoom users and guests.  & \\
2 & Webinar hosts can access AI-generated high-level summaries of webinar discussions, serving  & No \\
& $\qquad$ as an automated note-taking solution for both live and recorded sessions. Hosts can distribute  & \\
& $\qquad$ these summaries through the webinar's follow-up email workflow or their preferred  & \\
& $\qquad$ communication channels. Account owners and admins can control AI summary settings,    & \\
& $\qquad$ including usage permissions and notification preferences at both account and group levels. & \\
& \ldots $\qquad\qquad$ \ldots $\qquad\qquad$ \ldots $\qquad\qquad$ \ldots $\qquad\qquad$ \ldots $\qquad\qquad$ \ldots $\qquad\qquad$ \ldots $\qquad\qquad$ \ldots & \ldots \\
52 & The Cobrowse feature now supports web chat engagements, enabling agents to assist customers & No \\
& $\qquad$ more efficiently while maintaining privacy. When customers need help with website forms or  & \\
& $\qquad$ navigation, agents can request permission to view only the specific webpage where the issue   & \\
& $\qquad$ occurs. This enhancement allows for more precise, real-time support while protecting sensitive  & \\
& $\qquad$ information through built-in privacy safeguards. This feature must be enabled by Zoom. & \\
\hline
\end{tabular}}
\normalsize      
\vspace*{-0.50em}
\end{table*}

One instance from the Zoom dataset is shown in Table~\ref{table:OneInstance}. This instance has 71 total requirements: 19 were released in February 2025 and the other 52 were released in March 2025. The release timestamps give rise to the prioritization's ground truth in an authoritative way. When the 71 requirements of Table~\ref{table:OneInstance} are prioritized, the ground truth is to select the 19 requirements. Thus, we note the size of the ground truth here as $k$=19 and the size of the instance as $n$=71. This instance's size (71$>$50) illustrates the scalability challenge of manually prioritizing requirements~\cite{Perini-REW07}, echoing the importance of automatic solutions in industrial settings.

Our collection of user feedback data was done through a Google play store scraper~\cite{Google-Play-Scraper-Web}. Specifically, we extracted the reviews for the four software applications in the same time ranges as the release periods of the requirements. However, the reviews were collected one release cycle prior to the requirements, as shown in Table~\ref{table:Char}. For example, we extracted a total of 62,074 user reviews for Zoom from Jan 2022 to March 2025, but the Zoom requirements were collected from Feb 2022 to March 2025. This allowed us to ensure time sensitivity in evaluating ReFeed and iReFeed. In particular, we used the reviews up until the beginning of a prioritization instance. Additionally, to ensure relevant feedback, we used at most the two years of reviews preceding each prioritization instance. When fewer than two years of reviews were available, we used all available reviews. For example, we used the Zoom reviews from Jan 2023 to Jan 2025 for the instance shown in Table~\ref{table:OneInstance}, because the instance considered releasing the requirements in Feb 2025 and assuming the availability of the user reviews up till Jan 2025 was sensible to us.

Given the ground truth, we measure the qualities of ReFeed and iReFeed prioritization results by computing recall (R), precision (P), F$_1$-score (F$_1$), and F$_2$-score (F$_2$). These metrics quantify how much a prioritization solution overlaps the ground truth. Intuitively, recall indicates how complete the prioritization solution is whereas precision implies how noisy it is. Although F$_1$$=$$\frac{2 \cdot R \cdot P}{R + P}$ symmetrically represents both recall and precision in one metric, Berry \emph{et al.}~\cite{Berry-RE17} argue that, for automated tools supporting RE tasks in the context of large-scale software development, recall is more important than precision. Therefore, we also consider F$_2$$=$$\frac{5 \cdot R \cdot P}{4 \cdot P + R}$ which weighs recall twice as important as precision.

As shown in equations (\ref{eq:1})--(\ref{eq:3}), ReFeed and iReFeed rank candidate requirements in an ordered list according to requirements' priority scores. We thus investigate the qualities of the ordered list at five cutoff points: $k$, $k \pm (10\% \cdot n)$, and $k \pm (20\% \cdot n)$. Recall that $k$ is the ground truth size (e.g., $k$=19 in Table~\ref{table:OneInstance}) and $n$ is the instance size (e.g., $n$=71 in Table~\ref{table:OneInstance}). While top-$k$ represents a cutoff in experimental settings where the exact prioritization number is known, the others mimic practical scenarios. Taking Table~\ref{table:OneInstance}'s instance as an example, the selections of the top-ranked 5 (i.e., $19 - 20\% \cdot 71$) or 12 (i.e., $19 - 10\% \cdot 71$) requirements reflect aggressive prioritization decisions, whereas relaxed decisions are represented by choosing the top-ranked 26 (i.e., $19\,+\,10\% \cdot 71$) or 33 (i.e., $19\,+\,20\% \cdot 71$) requirements. 

To evaluate statistical significance, we performed the Wilcoxon signed-rank test across all cutoff points for each dataset. For each metric and iReFeed variant, we compared the scores against ReFeed. The resulting Wilcoxon p-values of the four iReFeed variants were then combined at each cutoff point using Fisher’s method.

\subsection{Results and Analysis}\label{subsect:42}

\begin{figure}[!t]
\centering
\includegraphics[width=\linewidth]{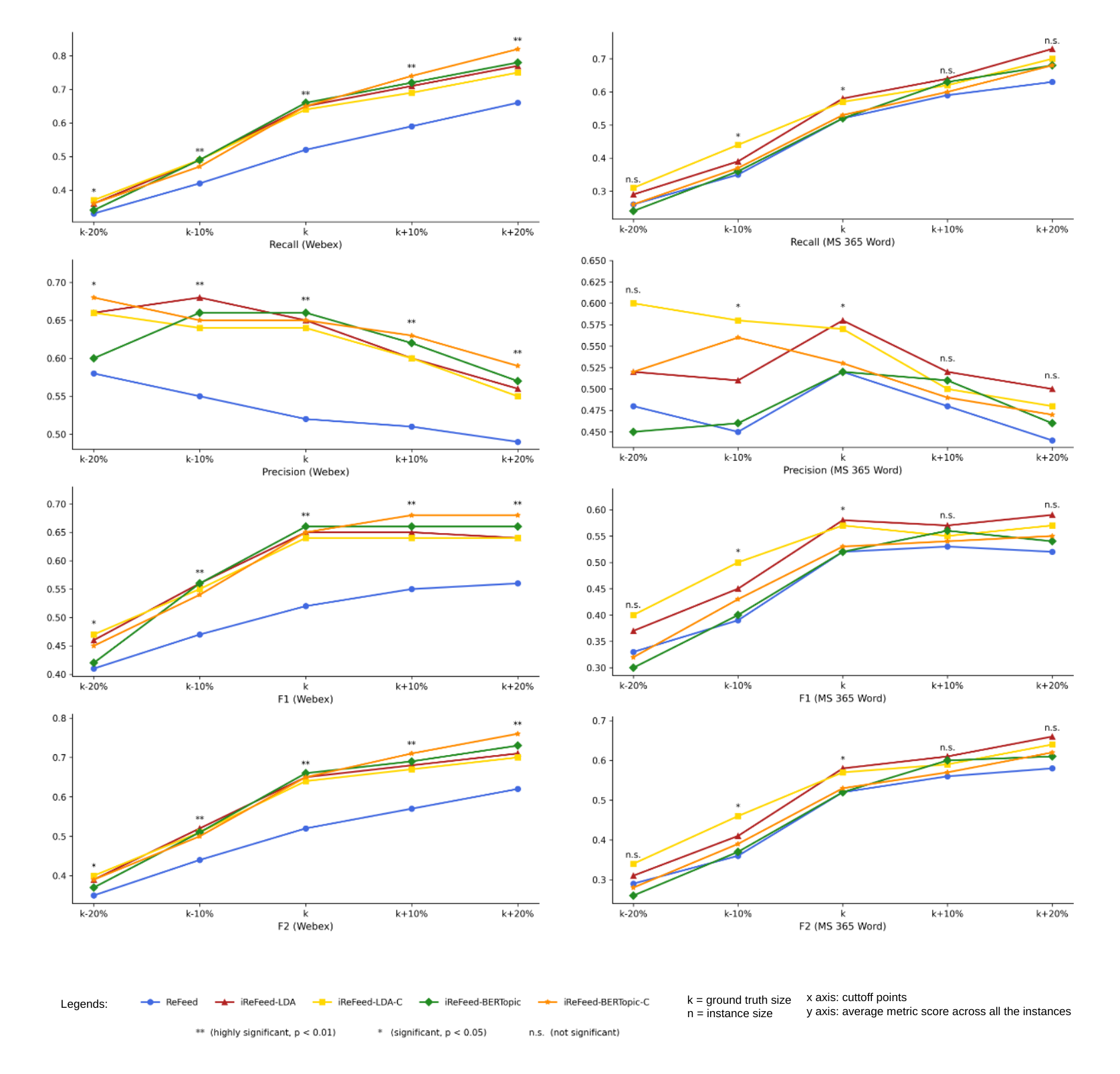}
\caption{Answering RQ$_1$ with average performances of ReFeed and the four variants of iReFeed: LDA, LDA-C, BERTopic, and BERTopic-C across Webex and MS 365 Word datasets.}
\label{fig:3a}
\end{figure}

\begin{figure}[!t]
\centering
\includegraphics[width=\linewidth]{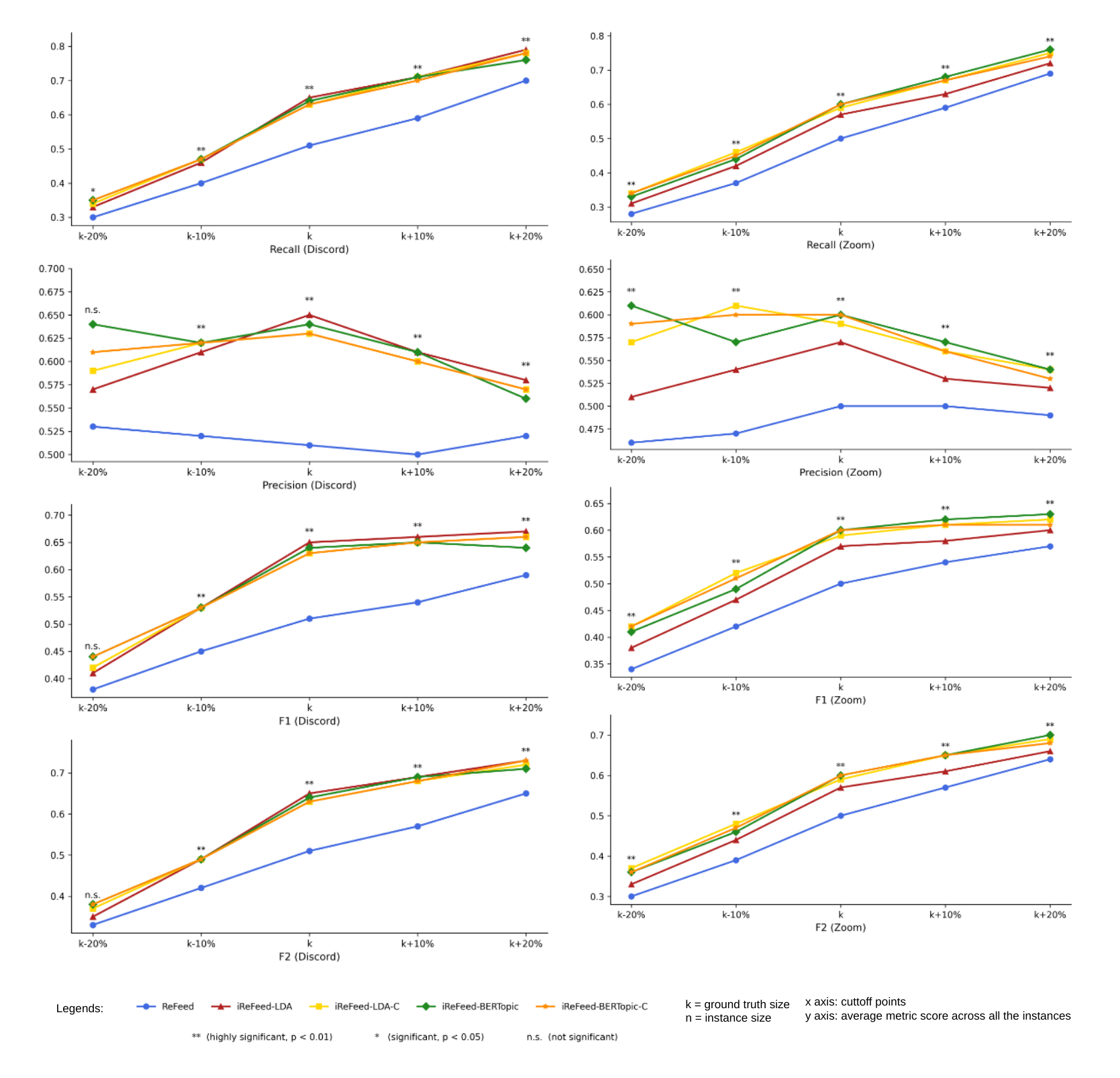}
\caption{Answering RQ$_1$ with average performances of ReFeed and the four variants of iReFeed: LDA, LDA-C, BERTopic, and BERTopic-C across Discord and Zoom datasets.}
\label{fig:3b}
\end{figure}

\begin{table}[]
\caption{F2 Performance of ReFeed and the four variants of iReFeed: LDA, LDA-C, BERTopic, and BERTopic-C across all the five cutoff points}\label{table:RQ1_results}
\begin{tabular}{|c|c|cccc|}
\hline
\multicolumn{6}{|c|}{\textbf{Webex}} \\ \hline
\multicolumn{1}{|l|}{} & \multicolumn{1}{|l|}{\multirow{2}{*}{\textbf{ReFeed}}} & \multicolumn{4}{|c|}{\textbf{iReFeed}} \\ \cline{3-6}
 & \multicolumn{1}{|l|}{} & \textbf{LDA} & \textbf{LDA-C} & \textbf{BERTopic} & \textbf{BERTopic-C} \\
 \hline
\textbf{k-20\%n} & 0.35 & 0.39 & \textbf{0.4} & 0.37 & 0.39 \\
\textbf{k-10\%n} & 0.44 & \textbf{0.52} & 0.51 & 0.51 & 0.50 \\
\textbf{k} & 0.52 & 0.65 & 0.64 & \textbf{0.66} & 0.65 \\
\textbf{k+10\%n} & 0.57 & 0.68 & 0.67 & 0.69 & \textbf{0.71} \\
\textbf{k+20\%n} & 0.62 & 0.71 & 0.70 & 0.73 & \textbf{0.76} \\
\hline
\multicolumn{6}{|c|}{\textbf{Microsoft 365 Word}} \\ \hline
\multicolumn{1}{|l|}{} & \multicolumn{1}{|l|}{\multirow{2}{*}{\textbf{ReFeed}}} & \multicolumn{4}{|c|}{\textbf{iReFeed}} \\ \cline{3-6}
 & \multicolumn{1}{|l|}{} & \textbf{LDA} & \textbf{LDA-C} & \textbf{BERTopic} & \textbf{BERTopic-C} \\
 \hline
\textbf{k-20\%n} & 0.29 & 0.31 & \textbf{0.34} & 0.26 & 0.28 \\
\textbf{k-10\%n} & 0.36 & 0.41 & \textbf{0.46} & 0.37 & 0.39 \\
\textbf{k} & 0.52 & \textbf{0.58} & 0.57 & 0.52 & 0.53 \\
\textbf{k+10\%n} & 0.56 & \textbf{0.61} & 0.59 & 0.60 & 0.57 \\
\textbf{k+20\%n} & 0.58 & \textbf{0.66} & 0.64 & 0.61 & 0.62 \\
\hline
\multicolumn{6}{|c|}{\textbf{Zoom}} \\ \hline
\multicolumn{1}{|l|}{} & \multicolumn{1}{|l|}{\multirow{2}{*}{\textbf{ReFeed}}} & \multicolumn{4}{|c|}{\textbf{iReFeed}} \\ \cline{3-6}
 & \multicolumn{1}{|l|}{} & \textbf{LDA} & \textbf{LDA-C} & \textbf{BERTopic} & \textbf{BERTopic-C} \\
 \hline
\textbf{k-20\%n} & 0.30 & 0.33 & \textbf{0.37} & 0.36 & 0.36 \\
\textbf{k-10\%n} & 0.39 & 0.44 & \textbf{0.48} & 0.46 & 0.47 \\
\textbf{k} & 0.50 & 0.57 & 0.59 & \textbf{0.60} & \textbf{0.60} \\
\textbf{k+10\%n} & 0.57 & 0.61 & \textbf{0.65} & \textbf{0.65} & \textbf{0.65} \\
\textbf{k+20\%n} & 0.64 & 0.66 & 0.69 & \textbf{0.70} & 0.68 \\
\hline
\multicolumn{6}{|c|}{\textbf{Discord}} \\ \hline
\multicolumn{1}{|l|}{} & \multicolumn{1}{|l|}{\multirow{2}{*}{\textbf{ReFeed}}} & \multicolumn{4}{|c|}{\textbf{iReFeed}} \\ \cline{3-6}
 & \multicolumn{1}{|l|}{} & \textbf{LDA} & \textbf{LDA-C} & \textbf{BERTopic} & \textbf{BERTopic-C} \\
 \hline
\textbf{k-20\%n} & 0.33 & 0.35 & 0.37 & \textbf{0.38} & \textbf{0.38} \\
\textbf{k-10\%n} & 0.42 & \textbf{0.49} & \textbf{0.49} & \textbf{0.49} & \textbf{0.49} \\
\textbf{k} & 0.51 & \textbf{0.65} & 0.63 & 0.64 & 0.63 \\
\textbf{k+10\%n} & 0.57 & \textbf{0.69} & 0.68 & \textbf{0.69} & 0.68 \\
\textbf{k+20\%n} & 0.65 & \textbf{0.73} & 0.72 & 0.71 & \textbf{0.73} \\
\hline
\end{tabular}
\end{table}

We plot the performances of ReFeed and iReFeed in Fig.~\ref{fig:3a} and Fig.~\ref{fig:3b}. The plots are organized horizontally by the four performance evaluation metrics, and vertically by the datasets. Each sub-figure shows the averages across the prioritization instances in that dataset. The combined significance levels are also reported in Fig.~\ref{fig:3a} and Fig.~\ref{fig:3b}. Table~\ref{table:RQ1_results} represents the F2 scores across all the datasets and the five cutoff points. Overall, a clear trend observed in Fig.~\ref{fig:3a} and Fig.~\ref{fig:3b} is that iReFeed consistently outperforms ReFeed. This shows that clustering requirements into topically coherent groups positively influences prioritization, compared with treating requirements as being independent from each other in the prioritization process.

Using ReFeed as the baseline, we notice from Fig.~\ref{fig:3a} and Fig.~\ref{fig:3b} that the improvements of recall are more prominent in Discord, and the precision improvements are more salient in Discord and Webex. We speculate a possible reason might be due to the user feedback's accurate matching and broad coverage of the candidate requirements in these datasets; however, testing the hypothesis requires future work. Nevertheless, a general trend of Fig.~\ref{fig:3a} and Fig.~\ref{fig:3b} is that LDA-C performs better than LDA, and BERTopic-C performs better than BERTopic. Therefore, our results suggest that integrating cluster's internal coherence into ranking requirements further enhances prioritization qualities.

When paying attention to different cutoff points, we identify the precision changes in Fig.~\ref{fig:3a} and Fig.~\ref{fig:3b}, as encouraging for iReFeed. As more top-ranked requirements are analyzed, it is not surprising that recall keeps increasing. For iReFeed, precision peaks at top-$k$ in most cases. If the prioritization decision is to continue including more requirements, then iReFeed drops precision more than ReFeed's precision decreases. This shows that the user feedback's influences on prioritization are more accurate to the top-$k$ ranked requirements than to further ranked requirements. In other words, iReFeed achieves a good and balanced performance when the number of prioritized requirements is near the ground-truth size $k$.

Taking the cutoff at $k$ and the weighted F$_2$ as a performance indicator, we note that LDA-C performs better than BERTopic-C in Microsoft 365 Word. LDA-C's performances in Discord, Webex, and Zoom are comparable to BERTopic-C's. Thus, among the four variants of iReFeed, we recommend LDA-C, though it is somewhat surprising that the pre-trained BERT model with extensive external data does not definitively outperform a locally operated LDA in user-feedback driven requirements prioritization. Through the above analyses, we conclude that iReFeed consistently outperforms ReFeed. We provide an example below to shed light on the differences between ReFeed and iReFeed. Consider two requirements and three feedback messages of a Microsoft 365 Word instance.

\begin{itemize}
\item r$_1$: ``Use @mentions in comments to let co-workers know when you need their input.''
\item r$_2$: ``Tired of being locked out of your document with macros? Now your docm files on OneDrive for Business allow simultaneous editing by multiple authors.''
\item m$_1$: ``This app is super easy for reading and sharing word files. It has all the important features required for writing but steps for using it should be mentioned or outline of all the editing options should be mentioned for beginners. For me, it saved a lot of time. Good app. Thank you''
\item m$_2$: ``Works great with Office 2016 and OneDrive across multiple devices/opearting systems. Can't do without it.''
\item m$_3$: ``I absolutely love Microsoft's Word app. I'm able to do so many important things like create documents that are truly professional, use the many templates to make the perfect letter or resume, upload document files in Word format even from the internet and edit, and if unlocked files I can highlight text and add notes. It allows me to be the author of my own document creations and either lock it, share it via other applications like Facebook or Twitter, and gives me the option to allow for others to participate in the writing and editing, for instance, a petition, and so much more.''
\end{itemize}

\begin{figure}[!t]
\centering
\includegraphics[width=0.80\columnwidth]{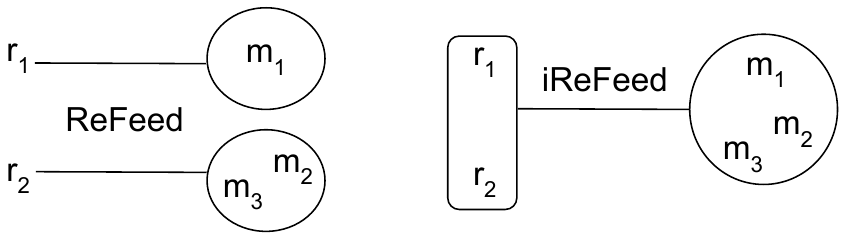}\\
\caption{Examples from a Microsoft 365 Word instance illustrating the differences between ReFeed (left) and iReFeed (right).}
\label{fig:4}
\end{figure}

Fig.~\ref{fig:4} depicts the associations established in ReFeed and iReFeed for the above entities. Although lexical clues can be directly spotted, e.g., ``mention'' between r$_1$ and m$_1$, iReFeed is able to recognize deeper semantic connections. For example, m$_1$ touches upon editing options and hence is linked to the simultaneous editing of r$_2$. Similarly, m$_3$ advocates collaborative writing which is what r$_1$ tries to facilitate. In fact, the cluster containing r$_1$ and r$_2$ resulted from iReFeed renders collaborative editing options as a topic. Consequently, iReFeed establishes richer associations that are missed in ReFeed.

\section{Uncovering ``Requires'' Pairs}\label{sect:5}

Our {\bf RQ$_2$ examines: ``To what extent is iReFeed conducive to identifying the `requires' relations?''} As discussed in Section~\ref{subsect:22}, ``requires'' relations greatly shape prioritization decisions. However, identifying such relations remains largely a manual task~\cite{Carlshamre-RE01, Aydemir-RE18}. With the advent of LLMs, researchers have exploited them to automate a wide variety of RE tasks. Specifically, LLMs like OpenAI's ChatGPT have shown promises in detecting the \emph{relationships} among requirements: inconsistency (e.g., ``the machine shall always offer coffee'' is in tension with ``the machine shall not offer Cappuccino as a beverage option'')~\cite{Fantechi-RE23}, traceability (e.g., an ``error decoding'' requirement is traced to a ``control and monitoring'' feature)~\cite{Rodriguez-REW23, Preda-MSR24}, satisfaction (e.g., whether a mobile app's specification satisfies GDPR's data withdrawal guidance)~\cite{Santos-RE24}, just to name a few.

Motivated by this thrust of research, we are interested in LLMs' capabilities of automatically uncovering ``requires'' relations for the requirements prioritization task. We chose the Word Processor benchmark to experiment because the 65 ``$r_a \to r_b$'' pairs defined on top of the 50 requirements~\cite{WP-Data-Web} would serve as the ground truth. Some examples from the Word Processor benchmark are: ``save a file'' $\to$ ``create a new file'' and ``redo the most recent change'' $\to$ ``undo a task and go back to previous state''. In terms of LLMs, we tested ChatGPT 4.5 and ChatGPT 4o because these are the latest models from OpenAI, though ChatGPT 4.5 is generally considered an improvement over ChatGPT 4o with improved reasoning and reduced hallucination rates ~\cite{Mori-GPT45-Web}. We developed a direct prompt and ran it through the LLMs' web interfaces in May 2025: 
\textit{A “requires” relation between two requirements
(req\_x and req\_y) is defined as: req x requires
req y for the purpose of software release, but
not vice versa. Identify and output all the
“requires” pairs from the requirements provided
below, using the format: req\_x --$>$ req\_y.
\{requirements\}}

In this prompt the objective of ``software release'' is made explicit, the instruction of ``identifying all the `requires' pairs'' is given, and the output format of ``req\_x $\to$ req\_y'' is specified. The set of requirements included in the prompt is denoted by \{requirements\}. 

Identifying requires relations involves understanding the context of different requirements. Although LLMs such as ChatGPT have large context windows, analyzing all requirements at once can still confuse the model with too many competing semantic cues, especially when the requirements are topically diverse. This motivates us to utilize clustering to improve focus by grouping requirements that are topically similar based on the user feedback, allowing the LLM to reason more effectively within a focused group of interconnected requirements. 
We fed all the 50 Word Processor \{requirements\} once as the {\bf baseline}. The {\bf iReFeed} prompt, however, would feed only the \{requirements\} of one cluster at a time. To produce the clusters for the Word Processor requirements, we first applied LDA-C to 146,310 reviews (Jan 2018--March 2025) of Microsoft 365 Word, as the answers to RQ$_1$ showed that LDA-C was among the best to implement iReFeed. We then used the user-review topics to group the 50 Word Processor requirements into clusters. This resulted in 7 clusters, and hence we ran the {\bf iReFeed} prompt 7 times, each time with the \{requirements\} from one and only one cluster.

We made sure to clear ChatGPT's history while prompting. Therefore, the {\bf baseline} and {\bf iReFeed} promptings were independent from each other, and so were the 7 times that the {\bf iReFeed} prompts were executed. Finally, we treated {\bf iReFeed} as a supplement to the {\bf baseline} prompting rather than its replacement. The reason was that {\bf iReFeed} prompt did not cross cluster boundaries, so the ``requires'' of two requirements in different clusters would not be detected by the {\bf iReFeed} prompts. Out of the 65 ground-truth ``requires'' relations, 29 were intra-cluster pairs, where both the requirements appeared within the same \textbf{iReFeed} cluster. Both ChatGPT 4.5 and ChatGPT 4o identified 7 of the 29 intra-cluster ``requires'' relations. The remaining 36 ground truth ``requires'' relations were inter-cluster relations and therefore could not be identified by the \textbf{iReFeed} prompts. We aggregated {\bf iReFeed} prompts' results into the {\bf baseline} results by eliminating duplicates, and referred to the aggregated ``requires'' pairs as the {\bf combined} results.

\begin{table}[!t]
\centering
\caption{Answering RQ$_2$ with Automatically Identified ``Requires'' Pairs for the Word Processor Benchmark~\cite{WP-Data-Web}}\label{table:RQ2}
\small
\begin{tabular}{ | c c | c c c c c |}
\hline
\multirow{2}{*}{LLM} & \multirow{2}{*}{prompting} & \# of dis- & \multirow{2}{*}{R} & \multirow{2}{*}{P} & \multirow{2}{*}{F$_1$} & \multirow{2}{*}{F$_2$} \\
& & tinct pairs & & & & \\
\hline \hline
Chat- & baseline & 18 & 0.08 & 0.28 & 0.12 & 0.09 \\
GPT & iReFeed & 26 & 0.11 &  0.27 &  0.15 & 0.12 \\
4.5 & combined & 37 & {\bf 0.17} & {\bf 0.30} & {\bf 0.22} & {\bf 0.19} \\
\hline
Chat- & baseline & 32 & 0.08 & 0.16 & 0.10 & 0.09 \\
GPT & iReFeed & 32 & 0.11 & 0.22 & 0.14 & 0.12 \\
4o & combined & 56 & {\bf 0.17} & 0.20 & 0.18 & 0.17 \\
\hline
\end{tabular}
\normalsize
\end{table}

The results' accuracies are summarized in Table~\ref{table:RQ2} where the best performances are displayed in boldface. While the complete results are shared in \url{https://doi.org/10.5281/zenodo.18881986}, we make a couple of observations. First, ChatGPT 4.5 generally outperforms ChatGPT 4o, which is not surprising given the advanced reasoning capabilities of ChatGPT 4.5~\cite{Mori-GPT45-Web}. Second, and more importantly, regardless of the LLMs, {\bf iReFeed} does help uncover additional ``requires'' pairs that the {\bf baseline} prompting fails to identify. This is encouraging in that, by zooming in only the requirements inside a single cluster, subtle relationships could be recognized. In a way, {\bf iReFeed} prompting decomposes a requirements set into smaller chunks, which slows down ChatGPT's reasoning and improves identification accuracies. We conclude that iReFeed adds value in ChatGPT's auto-generation of ``requires'' pairs.

Admittedly, the overall accuracies of Table~\ref{table:RQ2} are low. On one hand, interrelating requirements for prioritization may be a difficult task that requires LLM fine-tuning or advanced prompting like few-shot and chain-of-thought. On the other hand, the data leakage concerns appear to be alleviated. Data leakage means the LLM merely memorizes the data instead of doing the reasoning. Given that the Word Processor dataset~\cite{WP-Data-Web} was released in 2016 and ChatGPT 4.5 and 4o were released in February 2025 and May 2024 respectively, little leakage seemed to have occurred in our RQ$_2$ experiments.

\begin{table}[]
\centering
\caption{Additional baselines for RQ2 evaluation}\label{table:RQ2_baselines}
\begin{tabular}{|c|ccccc|}
\hline
\textbf{Baselines} & \multicolumn{1}{c}{\textbf{\# pairs}} & \multicolumn{1}{c}{\textbf{Precision}} & \multicolumn{1}{c}{\textbf{Recall}} & \multicolumn{1}{c}{\textbf{F1}} & \multicolumn{1}{c|}{\textbf{F2}} \\
\hline
\textbf{Random@18} & 18 & 0.03 & 0.01 & 0.01 & 0.01 \\
\textbf{Random@26} & 26 & 0.02 & 0.01 & 0.01 & 0.01 \\
\textbf{Random@32} & 32 & 0.03 & 0.01 & 0.02 & 0.01 \\
\textbf{Random@37} & 37 & 0.03 & 0.02 & 0.02 & 0.02 \\
\textbf{Random@56} & 56 & 0.03 & 0.02 & 0.02 & 0.02 \\
\hline
\textbf{TF-IDF + cosine} & 284 & 0.06 & 0.26 & 0.1 & 0.16 \\
\hline
\end{tabular}
\end{table}

To address the overall low accuracies reported in Table~\ref{table:RQ2}, we introduce additional baselines for comparison against the LLM-based results. First, we introduce random baselines. We construct the set of all ordered requirement pairs among the 50 requirements excluding self-pairs. This results in 50 x 49 = 2,450 possible pairs. The random baseline then samples directed requirement pairs from this set. As the evaluation results are affected by the number of predicted ``requires'' pairs, the number of randomly sampled pairs is matched with the number of pairs produced by each LLM setting in Table~\ref{table:RQ2}. Each random baseline is evaluated across 100 runs using different random samples for each run. The mean scores for precision, recall, F1, and F2 for the random variants are reported in Table~\ref{table:RQ2_baselines}. 

Wang et al. \cite{Wang_TSE_22} utilize TF-IDF and cosine similarity with a threshold of 0.125 for finding dependencies among requirements. Motivated by them, we introduce a TF-IDF $+$ cosine baseline in Table~\ref{table:RQ2_baselines}. We compute the pairwise TF-IDF and cosine similarity scores across all the 50 requirements and then utilize a threshold of 0.125 to identify the ``requires'' pairs. Since TF-IDF does not identify the direction of a ``requires'' relation, we convert each above threshold requirement pair (r\_i, r\_j) into directed candidates r\_i --$>$ r\_j and r\_j --$>$ r\_i. The candidate pairs are then evaluated against the ground truth ``requires” pairs. 

Table~\ref{table:RQ2_baselines} results indicate that the random baselines perform worse than the LLM variants,
indicating that ChatGPT and iReFeed assisted prompting performs better than random guessing. Additionally, while the TF-IDF and cosine baseline shows promise in terms of recall and F2, it struggles to maintain precision because it generates a large number of candidates. Overall, these results further indicate that automatic identification of ``requires'' relations is a challenging task.

\section{Integrating $\mathrm{iReFeed}$ into SBSE}\label{sect:6}

We explore in {\bf RQ$_3$: ``How could iReFeed be integrated into SBSE approaches to solving the NRP?''} To that end, we illustrate in this section a concrete way to fuse the ``requires'' pairs uncovered by iReFeed with NSGA-II, a well-suited SBSE solution to the NRP. Our illustration builds on the Word Processor dataset~\cite{WP-Data-Web}. In particular, we take the 50 requirements and use NSGA-II to search for solutions for a bi-objective optimization problem: Maximize the weighted sum of all the stakeholders' value scores (cf. columns 2--5 of Table~\ref{table:WP}) while at the same time minimizing the development resource estimates (i.e., column 8 of Table~\ref{table:WP}). Each solution produced by NSGA-II represents a possible release. A solution specifies which of the 50 candidate requirements are selected for release. Therefore, the possible solutions includes all possible subsets of the 50 candidate requirements. 

Our baseline implementation of NSGA-II follows the same algorithmic tuning as introduced by Finkelstein \emph{et al.}~\cite{Finkelstein-REJ09}: 

\begin{itemize}
\item The initial population is set to 200.
\item The experimental execution of NSGA-II is terminated after 50 generations, i.e., after 10,000 evaluations.
\item The genetic approach uses the tournament selection (with tournament size of 5), single-point crossover, and bitwise mutation.
\item The probability of the crossover operator being applied is set to $P_c$=0.8, and the probability of the mutation operator is set to $P_m$=$\frac{1}{\mathrm{\#\;of\;requirements}}$ =$\frac{1}{50}$=0.02.
\end{itemize}

\begin{figure}[!b]
\vspace*{-1.00em}
\centering
\includegraphics[width=1.00\columnwidth]{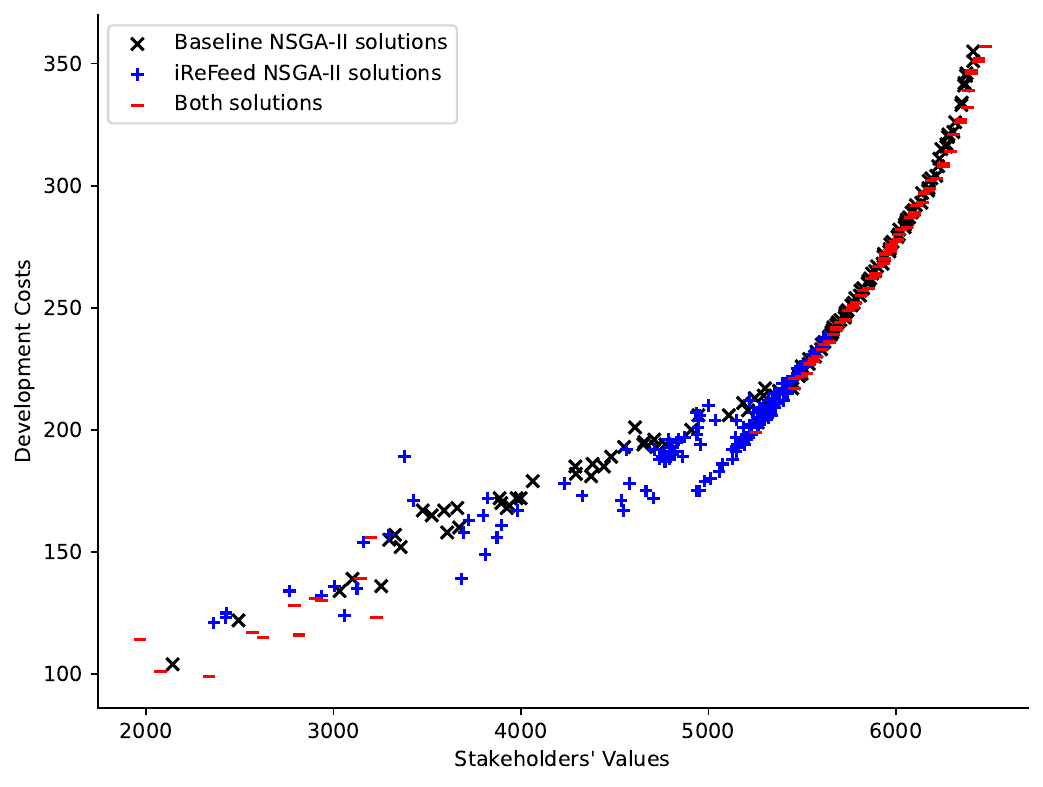}\\
\caption{One run of RQ$_3$ with search results from the baseline NSGA-II and the iReFeed D-value variant NSGA-II.}
\label{fig:5}
\vspace*{-1.00em}
\end{figure}

\vspace*{+0.50em}

The result of such a search is a Pareto front. Each element on this front is a candidate solution to the NRP. All solutions on the Pareto front are non-dominated: no other solution on the front is better according to both objectives. The Pareto front thus represents a set of ``best compromises'' between the objectives that can be found by the search-based algorithm. As an illustration, we show the search results of the baseline NSGA-II run in Fig.~\ref{fig:5} in which no ``requires'' pairs are taken into account. In this figure, the ``$\times$'' and ``$-$'' nodes denote the Pareto-optimal solutions found by the baseline NSGA-II algorithm.

To integrate iReFeed into NSGA-II, we introduce dependency value (D-value). We utilize the automatically identified requires pairs from ChatGPT 4.5 combined results in RQ2 to compute the D-value. Let $\mathcal{D}$ be the set of all the requires pairs identified by ChatGPT 4.5 in RQ2, where each pair ``$r_a \to r_b$'' indicates that requirement a requires requirement b. Let $count_i$ be defined as the total number of requires relations where i appears on the right hand side of a requires pair ($count_i$= |\{ ($r_a \to r_b$) $\in$ $\mathcal{D}$, where b=i\}|).
The D-value for a requirement i is defined as follows:

\[
D\text{-}value_i = \frac{count_i}{\mathcal{D}}
\]

\noindent D-value for iReFeed is incorporated as a third objective of maximizing D-value. We maximize D-value as an objective in iReFeed giving importance to the requirements with higher dependencies for selection. The D values have a short range and are skewed in nature, so we apply different transformations and normalization techniques to explore alternative representations of the dependency values. We introduce five variants: D-value, log transformed D value, power transformed D-value, z-score normalized D-value, and inverse D-value. The formulas for each of the variants are reported in Table \ref{table:D_value_variants}. For the log transformation, we apply $\log_e(1 + D_i)$ to avoid undefined values when $D_i=0$. When applying the z-score normalization, we shift each value by the magnitude of the most negative z-score to ensure that all the values are non-negative, while preserving the relative ordering between the values. The inverse D-value serves as a contrasting variant that prioritizes requirements with fewer dependencies. This enables us to evaluate how the search is affected by the dependency values.

\begin{table}[t]
\centering
\caption{Dependency Value (D-value) Variants}
\label{table:D_value_variants}
\begin{tabular}{|c|c|}
\hline
\textbf{Variant} & \textbf{Formula} \\
\hline
D-value ($D_i$)
& $\dfrac{count_i}{|\mathcal{D}|}$ \\
Log-transformed D-value 
& $\log_e(1 + D_i)$ \\
Inverse D-value 
& $ 1 - D_i$ \\
Power-transformed D-value 
& $ D_i^{0.5}$ \\
Z-score normalized D-value 
& $ \dfrac{D_i - \mu}{\sigma}$ \\
&\\
\hline
\end{tabular}
\end{table}

\begin{table}[]
\caption{RQ3: average across 30 runs of percentage of solutions in reference Pareto front}
\label{table:RQ3_average_runs}
\begin{tabular}{|c|cc|c|}
\hline
\textbf{Variants} & \textbf{\begin{tabular}[c]{@{}c@{}}Baseline \\ NSGA-II\end{tabular}} & \textbf{\begin{tabular}[c]{@{}c@{}}iReFeed \\ NSGA-II\end{tabular}} & \textbf{\begin{tabular}[c]{@{}c@{}}Vargha-Delaney \\ {$\mathbf{A_{12}}$}\end{tabular}} \\
\hline
D-value & 80.75 & \textbf{97.05} & 0.96 \\
Log transformed D-value & 82.56 & \textbf{96.31} & 0.95\\
Inverse D-value & \textbf{96.03} & 67.14 & 0\\
Power transformed D-value & 82.12 & \textbf{97.94} & 1.0 \\
Z-score normalized D-value & 83.80 & \textbf{96.85} & 0.97\\
\hline
\end{tabular}
\end{table}

\begin{figure}[!t]
\centering
\includegraphics[width=\linewidth]{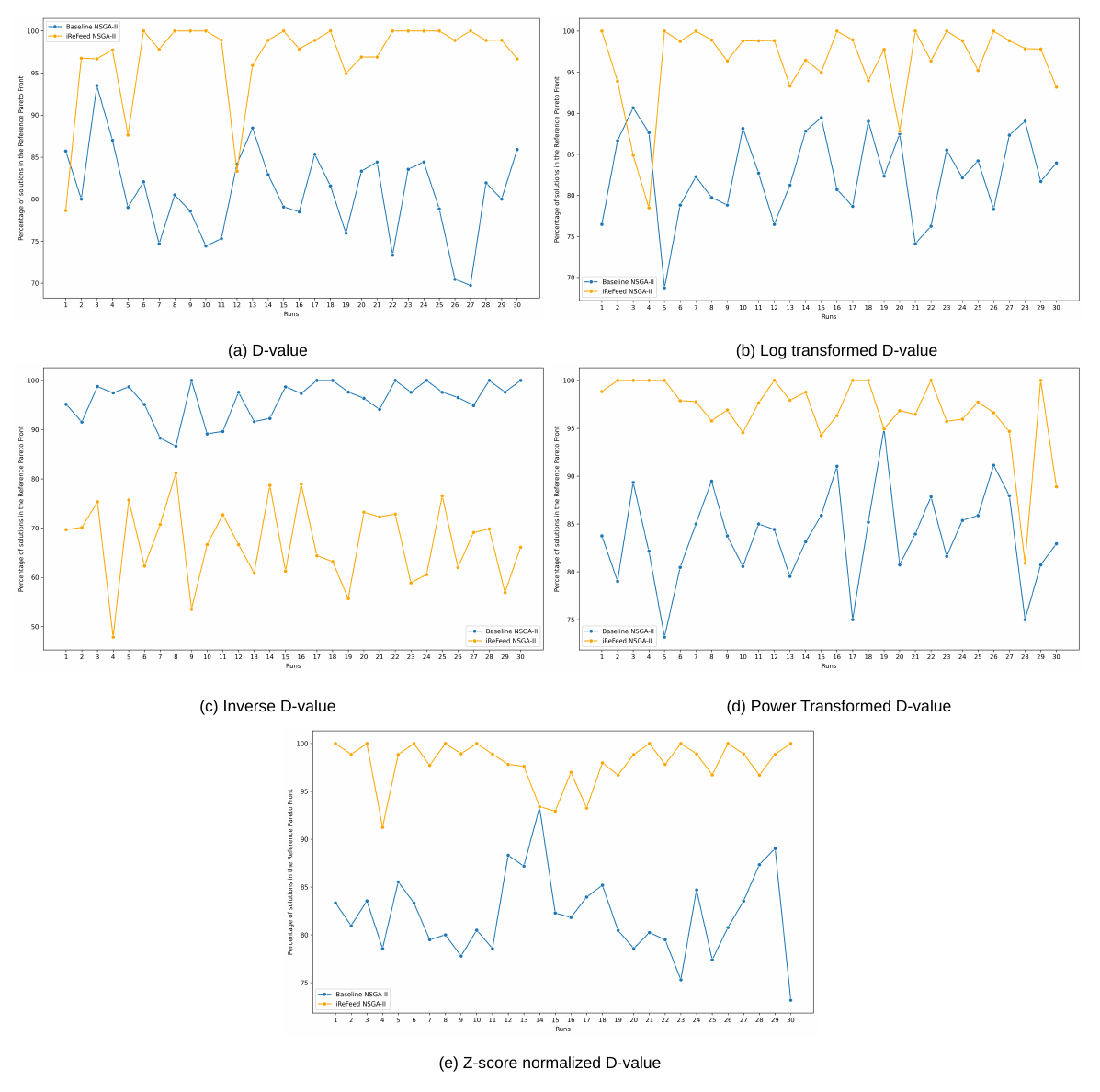}
\caption{RQ3: Comparison of each iReFeed variant against the baseline across 30 runs}
\label{fig:RQ3_plot}
\end{figure}

In Fig.~\ref{fig:5}, the ``$+$'' and ``$-$'' nodes denote the Pareto-optimal solutions found by the iReFeed NSGA-II search for one of the runs (run 8) of the D value variant . The shared solutions are marked by ``$-$''.
To evaluate the solutions found by baseline NSGA-II and iReFeed NSGA-II, we construct a \emph{reference Pareto front} informed by the work of Finkelstein \emph{et al.}~\cite{Finkelstein-REJ09}. In a nutshell, the reference Pareto front is the result of merging the solutions from different search algorithms and then excluding the solutions that are dominated by some other merged solutions. Since we are solving for the bi-objective optimization problem, we only consider the value and cost pairs as solutions when evaluating the baseline NSGA-II and iReFeed NSGA-II solutions.  In Fig.~\ref{fig:5}, for example, although the solution $A$: (value=5329, cost=209) is Pareto-optimal when only the baseline NSGA-II is considered, $A$ becomes dominated by one of the iReFeed NSGA-II solutions $B$: (value=5335, cost=206). The reason is that $B$'s value is higher than $A$'s value, and at the same time, $B$'s cost is lower than $A$'s cost. In the face of $B$, $A$ is no longer on the reference Pareto front. Therefore, the reference Pareto front denotes the best available approximation to the \emph{real} Pareto front~\cite{Finkelstein-REJ09}.

The measure used by Finkelstein \emph{et al.}~\cite{Finkelstein-REJ09} to compare different search algorithms is to count the number of individual algorithm's solutions on the reference Pareto front, namely
the solutions that are not dominated by the reference
Pareto front. We report the average results for each iReFeed variant against the baseline across 30 runs in Table \ref{table:RQ3_average_runs}. We also compute the Vargha--Delaney $A_{12}$ \cite{Vargha_Delaney} effect size across the 30 runs to measure the consistency of performance difference between the two approaches. An $A_{12}$ value greater than 0.5 indicates that iReFeed variant tends to achieve better performance, while an $A_{12}$ value less than 0.5 indicates that baseline tends to achieve better performance. Additionally, the results across each individual run for all the iReFeed variants against baseline are shown in Figure \ref{fig:RQ3_plot}. 

We observe from Table \ref{table:RQ3_average_runs} and Figure \ref{fig:RQ3_plot} that for all iReFeed variants except the inverse D-value variant, iReFeed search solutions represent a higher proportion of solutions in the reference Pareto front. The greater share, according to Finkelstein \emph{et al.}~\cite{Finkelstein-REJ09}, suggests the better performance of iReFeed NSGA-II compared to baseline NSGA-II. The Vargha--Delaney $A_{12}$ scores further support this observation. Apart from the inverse D-value variant, all other variants have $A_{12}$ values closer to 1.0 indicating that the iReFeed variant consistently across runs represents a higher percentage of solutions on the reference Pareto front compared to the baseline. In contrast, the $A_{12}$ value of the inverse D-value variant is 0 indicating that the baseline consistently outperforms the iReFeed variant, when the importance of D-value is reversed. Thus, we conclude positive findings of RQ$_3$ with iReFeed's superiority over the baseline SBSE solution. We attribute this to the baseline's obliviousness to requirements interconnectedness.

The inverse D-value variant is particularly interesting. By inverting the importance of dependency values, this variant prioritizes requirements with less dependency on other requirements. As a result, its performance drops significantly compared to baseline. This variant further strengthens the importance of integrating D-value into the iReFeed search. 

\section{Threats to Validity}\label{sect:7a}
The core constructs of our experiments for RQ1 are requirements prioritization instance and solution quality. Each instance contains the requirements that shall be prioritized to release and those that shall not. The software projects' own release histories allow for the ground truth to be defined authoritatively. Although the instance can be easily extended to a series of multiple releases, we are interested in only two consecutive release periods, as this is also the focus of ReFeed. We utilize user feedback for prioritization. However, the actual release order of requirements may also be influenced by additional factors such as technical dependencies, implementation effort, product strategy, regulatory issues, and internal roadmaps. Moreover, our prioritization instance is based on requirements from two consecutive release instances, as opposed to selection of requirements from an open backlog. We measure the prioritization solution's quality with well-known metrics at various cutoff points. Another threat involves the intention score used for prioritization. The highest intention score is assigned for feedback associated with feature request. Therefore, feedback conveying other intentions such as bug fixes and non functional concerns may receive lower scores for priority, even though such concerns may be an important factor for release planning.

The internal validity of the RQ1 experiments could be affected by our choice of extracting review data with a one-release-cycle buffer ahead of requirements data. This avoids solving a prioritization instance without any review data. A related decision is our use of cumulative reviews, rather than selecting only recent ones. While our rationale is to maintain a broad time range to capture both short-term and long-term trends of requirements evolution and integration of user feedback, determining the optimal amount of review data to exploit in prioritization is beyond the scope of our current work. Keep in mind that, for each prioritization instance, our experiments of ReFeed and iReFeed have used the same amounts of review and requirements data. As a result, the comparisons reported in RQ1 reflect the differences between ReFeed and iReFeed.

The RQ1 results are based on the four real-world datasets that we have curated. Due to the threats to external validity, it would be too strong to claim iReFeed's better performance over ReFeed on other prioritization instances (e.g., the software applications outside the video conferencing, word processing, and social networking domains). In terms of transparency and reliability, we share our datasets, along with our implementations and experimental results, in \url{https://doi.org/10.5281/zenodo.18881986} to facilitate replication and expansion.

Our study for RQ3 has several limitations. From the construct validity perspective, we use an individual search algorithm's solutions shared within the reference Pareto front to evaluate that algorithm's performance. While Finkelstein \emph{et al.}~\cite{Finkelstein-REJ09} have used the same evaluative construct, there exist other measures, such as convergence and hypervolume~\cite{Zhang-TOSEM18}, which could be considered in future assessments. We believe the internal validity is high, as the search results reported for RQ3 are based on the same parameter setting for both the baseline and the iReFeed NSGA-II algorithms. Thus, the different shares of the reference Pareto front must be caused by the ``requires'' pairs exploited in iReFeed NSGA-II. Our results may not generalize to other datasets or SBSE algorithms---a threat to the external validity. With the encouraging initial results obtained for RQ3, we are positive that iReFeed is amenable to be integrated into other metaheuristic search techniques like the two-archive algorithm~\cite{Finkelstein-REJ09} or even hyper-heuristic search methods~\cite{Zhang-TOSEM18}.

\section{Conclusion}\label{sect:7}

Modern software engineering often injects agile's iterative and incremental development. Requirements are continuously delivered, though challenges exist in terms of lack of long-term planning, poor resource management, and scope creep~\cite{Raymond-Web}. User feedback provides an invaluable source for supporting agile software project's strategic releases. In this paper, we have presented a novel approach to enhancing user-feedback driven requirements prioritization. Our main novelty lies in the clustering of requirements based on the topics that emerged from user reviews. Our evaluations show that iReFeed not only outperforms ReFeed in making prioritization decisions, but also is conducive to enabling ChatGPT to identify requirements dependencies. The integration of iReFeed and NSGA-II, to the best of our knowledge, is the first to synthesize CrowdRE and SBSE for requirements prioritization.

Our future work includes carrying out experimentation on more datasets, investigating the optimal amount of feedback data to use, testing advanced prompting methods like few-shot and chain-of-thought, and guiding metaheuristic or hyper-heuristic search proactively with the ``requires'' pairs.

\bibliographystyle{elsarticle-num}
\bibliography{references}

@Book{ruhe10,
  author = {G{\"{u}}nther Ruhe},
  title = {Product Release Planning - Methods, Tools and Applications},
  publisher = {CRC Press},
  year = {2010},
}

@Book{leffingwell03,
  author = {Dean Leffingwell and Don Widrig},
  title = {Managing Software Requirements: A Use Case Approach},
  publisher = {Addison-Wesley},
  year = {2003},
  url = {https://dl.acm.org/doi/10.5555/829554}
}

@article{Qi-arXiv20,
	title        = {Stanza: A {Python} Natural Language Processing Toolkit for Many Human Languages},
	author       = {Peng Qi and Yuhao Zhang and Yuhui Zhang and Jason Bolton and Christopher D. Manning},
	year         = {2020},
	month        = {March},
	journal      = {CoRR},
	url = {https://doi.org/10.48550/arXiv.2003.07082},
}

@article{Durillo-ESE11,
	title        = {A study of the bi-objective next release problem},
	author       = {Juan Jos{\'{e}} Durillo and Yuanyuan Zhang and Enrique Alba and Mark Harman and Antonio J. Nebro},
	year         = 2011,
	month        = {February},
	journal      = {Empirical Software Engineering},
	volume       = 16,
	number       = 1,
	pages        = {29--60},
	url          = {https://doi.org/10.1007/s10664-010-9147-3}
}

@article{Silva-ESE21,
	title        = {Topic modeling in software engineering research},
	author       = {Camila Mariane C. Silva and Matthias Galster and Fabian Gilson},
	year         = 2021,
	month        = {November},
	journal      = {Empirical Software Engineering},
	volume       = 26,
	number       = 6,
	pages        = {120:1--120:62},
	url          = {https://doi.org/10.1007/s10664-021-10026-0}
}

@article{Zhang-TOSEM18,
  author = {Yuanyuan Zhang and Mark Harman and Gabriela Ochoa and G{\"{u}}nther Ruhe and Sjaak Brinkkemper},
  title = {An Empirical Study of Meta- and Hyper-Heuristic Search for Multi-Objective Release Planning},
  journal = {ACM Transactions on Software Engineering and Methodology},
  volume = {27},
  number = {1},
  month = {June},
  year = {2012},
  pages = {3:1--3:32},
  url = {https://doi.org/10.1145/3196831}
}

@article{Finkelstein-REJ09,
	title        = {A search based approach to fairness analysis in requirement assignments to aid negotiation, mediation and decision making},
	author       = {Anthony Finkelstein and Mark Harman and S. Afshin Mansouri and Jian Ren and Yuanyuan Zhang},
	year         = {2009},
	month        = {December},
	journal      = {Requirements Engineering},
	volume       = {14},
	number       = {4},
	pages        = {231--245},
	url          = {https://doi.org/10.1007/s00766-009-0075-y}
}

@article{Cai12,
  author = {Xinye Cai and Ou Wei and Zhiqiu Huang},
  title = {Evolutionary approaches for multi-objective next release problem},
  journal = {Computing and Informatics},
  volume = {31},
  number = {4},
  year = {2012},
  pages = {847--875},
}

@article{Rahimi-ASC23,
  author = {Iman Rahimi and Amir H. Gandomi and Mohammad Reza Nikoo and Fang Chen},
  title = {A comparative study on evolutionary multi-objective algorithms for next release problem},
  journal = {Applied Soft Computing},
  volume = {52},
  number = {3},
  month = {September},
  year = {2023},
  pages = {110472:1--110472:13},
  doi = {https://doi.org/10.1016/j.asoc.2023.110472}
}

@article{Karlsson-SW97,
  author = {Joachim Karlsson and Kevin Ryan},
  title = {A Cost-Value Approach for Prioritizing Requirements},
  journal = {IEEE Software},
  volume = {14},
  number = {5},
  month = {September/October},
  year = {1997},
  pages = {67--74},
  url = {https://doi.org/10.1109/52.605933}
}

@article{Spinellis-SW15,
  author = {Diomidis Spinellis},
  title = {The Strategic Importance of Release Engineering},
  journal = {IEEE Software},
  volume = {32},
  number = {2},
  month = {March/April},
  year = {2015},
  pages = {3--5},
  url = {https://doi.org/10.1109/MS.2015.54},
}

@article{Marner-Computers22,
  author = {Kristina Marner and Stefan Wagner and Guenther Ruhe},
  title = {Release Planning Patterns for the Automotive Domain},
  journal = {Computers},
  volume = {11},
  number = {6},
  month = {June},
  year = {2022},
  pages = {89:1--89:26},
  url = {https://doi.org/10.3390/computers11060089}
}

@article{Xuan-TSE12,
  author = {Jifeng Xuan and He Jiang and Zhilei Ren and Zhongxuan Luo},
  title = {Solving the Large Scale Next Release Problem with a Backbone-Based Multilevel Algorithm},
  journal = {IEEE Transactions on Software Engineering},
  volume = {38},
  number = {5},
  month = {September-October},
  year = {2012},
  pages = {261--284},
  url = {https://doi.org/10.1109/TSE.2011.92}
}

@article{Scalabrino-TSE19,
  author = {Simone Scalabrino and Gabriele Bavota and Barbara Russo and Massimiliano {Di Penta} and Rocco Oliveto},
  title = {Listening to the Crowd for the Release Planning of Mobile Apps},
  journal = {IEEE Transactions on Software Engineering},
  volume = {45},
  number = {1},
  month = {January},
  year = {2019},
  pages = {68--86},
  url = {https://doi.org/10.1109/TSE.2017.2759112}
}

@article{Perini-TSE13,
  author = {Anna Perini and Angelo Susi and Paolo Avesani},
  title = {A Machine Learning Approach to Software Requirements Prioritization},
  journal = {IEEE Transactions on Software Engineering},
  volume = {39},
  number = {4},
  month = {April},
  year = {2013},
  pages = {445--461},
  url = {https://doi.org/10.1109/TSE.2012.52}
}

@article{Karlsson-IST98,
  author = {Joachim Karlsson and Claes Wohlin and Bj{\"{o}}rn Regnell},
  title = {An evaluation of methods for prioritizing software requirements},
  journal = {Information \& Software Technology},
  volume = {39},
  number = {14-15},
  year = {1998},
  pages = {939--947},
  url = {https://doi.org/10.1016/S0950-5849(97)00053-0}
}

@article{Kifetew-IST21,
  author = {Fitsum Meshesha Kifetew and Anna Perini and Angelo Susi and Alberto Siena and Denisse Mu\~{n}ante and Itzel Morales-Ramirez},
  title = {Automating user-feedback driven requirements prioritization},
  journal = {Information \& Software Technology},
  volume = {138},
  month = {October},
  year = {2021},
  pages = {106635:1--106635:16},
  url = {https://doi.org/10.1016/j.infsof.2021.106635}
}

@article{Bagnall-IST01,
  author = {Anthony J. Bagnall and Victor J. Rayward-Smith and Ian M. Whittley},
  title = {The next release problem},
  journal = {Information \& Software Technology},
  volume = {43},
  number = {14},
  month = {December},
  year = {2001},
  pages = {883--890},
  url = {https://doi.org/10.1016/S0950-5849(01)00194-X}
}

@article{Svahnberg-IST10,
  author = {Mikael Svahnberg and Tony Gorschek and Robert Feldt and Richard Torkar and Saad Bin Saleem and Muhammad Usman Shafique},
  title = {A systematic review on strategic release planning models},
  journal = {Information \& Software Technology},
  volume = {52},
  number = {3},
  month = {March},
  year = {2010},
  pages = {237--248},
  url = {https://doi.org/10.1016/j.infsof.2009.11.006}
}

@misc{Raymond-Web,
  author = {Daniel Raymond},
  title = {{Top 10 Cons or Disadvantages of Agile Methodology}},
  year = {2023},
  howpublished = {\url{https://projectmanagers.net/top-10-cons-or-disadvantages-of-agile-methodology/}},
  note={{Last} accessed: \today}
}

@misc{Mori-GPT45-Web,
  author = {Giancarlo Mori},
  title = {{GPT-4.5 vs GPT-4o: Comparing OpenAI's Latest AI Models}},
  year = {2025},
  howpublished = {\url{https://giancarlomori.substack.com/p/gpt-45-vs-gpt-4o-comparing-openais}},
  note={{Last} accessed: \today}
}

@misc{Google-Play-Scraper-Web,
  author = {{Python Software Foundation}},
  title = {{Google-Play-Scraper}},
  year = {2026},
  howpublished = {\url{https://pypi.org/project/google-play-scraper/}},
  note={{Last} accessed: \today}
}

@misc{WP-Data-Web,
  author = {Muhammad Rezaul Karim and G{\"{u}}nther Ruhe},
  title = {Datasets of ``Bi-objective Genetic Search for Release Planning in Support of Themes''},
  year = {2016},
  howpublished = {\url{https://sites.google.com/site/mrkarim/data-sets}},
  note={{Last} accessed: \today}
}

@misc{ZoomReleaseNotes-Web,
  author = {{Zoom}},
  title = {{Release Notes for Windows}},
  year = {2026},
  howpublished = {\url{https://support.zoom.us/hc/en-us/articles/201361953-Release-notes-for-Windows}},
  note={{Last} accessed: \today}
}

@misc{Microsoft365ReleaseNotes-Web,
  author = {Microsoft},
  title = {{Microsoft 365 Apps for Windows: Archived Release Notes}},
  year = {2026},
  howpublished = {\url{https://learn.microsoft.com/en-us/officeupdates/monthly-channel-archived}},
  note={{Last} accessed: \today}
}

@misc{DiscordReleaseNotes-Web,
  author = {Discord},
  title = {{Change Log}},
  year = {2026},
  howpublished = {\url{https://discord.com/developers/docs/change-log}},
  note={{Last} accessed: \today}
}

@misc{WebexReleaseNotes-Web,
  author = {Cisco},
  title = {{Webex App: What's New}},
  year = {2026},
  howpublished = {\url{https://help.webex.com/en-us/article/8dmbcr/Webex-App-|-What's-New}},
  note={{Last} accessed: \today}
}

@inproceedings{Carlshamre-RE01,
	title        = {An industrial survey of requirements interdependencies in software product release planning},
	author       = {P{\"{a}}r Carlshamre and Kristian Sandahl and Mikael Lindvall and Bjorn Regnell and Johan Natt och Dag},
	year         = {2001},
	month        = {August},
	booktitle    = {Proceedings of the 5th IEEE International Symposium on Requirements Engineering (RE)},
	address      = {Toronto, Canada},
	pages        = {84--91},
	url          = {https://doi.org/10.1109/ISRE.2001.948547}
}

@inproceedings{Aydemir-RE18,
	title        = {The Next Release Problem Revisited: A New Avenue for Goal Models},
	author       = {Fatma Basak Aydemir and Fabiano Dalpiaz and Sjaak Brinkkemper and Paolo Giorgini and John Mylopoulos},
	year         = {2018},
	month        = {August},
	booktitle    = {Proceedings of the 26th IEEE International Requirements Engineering Conference (RE)},
	address      = {Banff, Canada},
	pages        = {5--16},
	url          = {https://doi.org/10.1109/RE.2018.00-56}
}

@inproceedings{Feather-RE02,
	title        = {Converging on the Optimal Attainment of Requirements},
	author       = {Martin S. Feather and Tim Menzies},
	year         = {2002},
	month        = {September},
	booktitle    = {Proceedings of the 10th IEEE Joint International Conference on Requirements Engineering (RE)},
	address      = {Essen, Germany},
	pages        = {263--272},
	url          = {https://doi.org/10.1109/ICRE.2002.1048537}
}

@inproceedings{Perini-REW07,
	title        = {An Empirical Study to Compare the Accuracy of AHP and CBRanking Techniques for Requirements Prioritization},
	author       = {Anna Perini and  Angelo Susi and Filippo Ricca and Cinzia Bazzanella},
	year         = {2007},
	month        = {October},
	booktitle    = {Proceedings of the 5th International Workshop on Comparative Evaluation in Requirements Engineering (CERE)},
	address      = {New Delhi, India},
	pages        = {23--35},
	url          = {https://doi.org/10.1109/CERE.2007.1}
}

@inproceedings{Rodriguez-REW23,
	title        = {Prompts Matter: Insights and Strategies for Prompt Engineering in Automated Software Traceability},
	author       = {Alberto D. Rodriguez and Katherine R. Dearstyne and Jane Cleland-Huang},
	year         = {2023},
	month        = {September},
	booktitle    = {Proceedings of the 11th International Workshop on Software and Systems Traceability (SST)},
	address      = {Hannover, Germany},
	pages        = {455--464},
	url          = {https://doi.org/10.1109/REW57809.2023.00087}
}

@inproceedings{Stronstad-REW23,
	title        = {What's Next in my Backlog? {T}ime Series Analysis of User Reviews},
	author       = {G{\o{}}ran H. Str{\o{}}nstad and Ilias Gerostathopoulos and Emitz{\'{a}} Guzm{\'{a}}n},
	year         = {2023},
	month        = {September},
	booktitle    = {Proceedings of the 8th International Workshop on Empirical Requirements Engineering (EmpiRE)},
	address      = {Hannover, Germany},
	pages        = {154--161},
	url          = {https://doi.org/10.1109/REW57809.2023.00032}
}

@Incollection{Nayebi15,
  author        = {Maleknaz Nayebi and Guenther Ruhe},
  editor        = {Christian Bird and Tim Menzies and Thomas Zimmermann},
  title         = {Analytical Product Release Planning},
  booktitle     = {The Art and Science of Analyzing Software Data},
  publisher     = {Morgan Kaufmann},
  year          = {2015},
  pages         = {555--589},
  url = {https://doi.org/10.1016/b978-0-12-411519-4.00019-7}
}

@inproceedings{Groen-REFSQ15,
  author = {Eduard C. Groen and J{\"{o}}rg D{\"{o}}rr and Sebastian Adam},
  title = {Towards Crowd-Based Requirements Engineering A Research Preview},
  booktitle = {Proceedings of the 21st International Working Conference on Requirements Engineering: Foundation for Software Quality (REFSQ)},
  pages = {247--253},
  month = {March},
  year = {2015},
  address = {Essen, Germany},
  url          = {https://doi.org/10.1007/978-3-319-16101-3_16}
}

@inproceedings{Zorn-Pauli-REFSQ13,
  author = {Gabriele Zorn-Pauli and Barbara Paech and Tobias Beck and Hannes Karey and G{\"{u}}nther Ruhe},
  title = {Analyzing an Industrial Strategic Release Planning Process - A Case Study at {R}oche {D}iagnostics},
  booktitle = {Proceedings of the 19th International Working Conference on Requirements Engineering: Foundation for Software Quality (REFSQ)},
  pages = {269--284},
  month = {April},
  year = {2013},
  address = {Essen, Germany},
  url          = {https://doi.org/10.1007/978-3-642-37422-7_19}
}

@inproceedings{Preda-MSR24,
author = {Preda, Anamaria-Roberta and Mayr-Dorn, Christoph and Mashkoor, Atif and Egyed, Alexander},
title = {Supporting High-Level to Low-Level Requirements Coverage Reviewing with Large Language Models},
year = {2024},
isbn = {9798400705878},
publisher = {Association for Computing Machinery},
address = {New York, NY, USA},
url = {https://doi.org/10.1145/3643991.3644922},
doi = {10.1145/3643991.3644922},
booktitle = {Proceedings of the 21st International Conference on Mining Software Repositories},
pages = {242–253},
numpages = {12},
location = {Lisbon, Portugal},
series = {MSR '24}
}

@inproceedings{Santos-RE24,
  author={Santos, Sarah and Breaux, Travis and Norton, Thomas and Haghighi, Sara and Ghanavati, Sepideh},
  booktitle={2024 IEEE 32nd International Requirements Engineering Conference (RE)}, 
  title={Requirements Satisfiability with In-Context Learning}, 
  year={2024},
  volume={},
  number={},
  pages={168-179},
  doi={10.1109/RE59067.2024.00025}
}

@inproceedings{Sihag-RE23,
  author = {Manish Sihag and Ze Shi Li and Amanda Dash and Nowshin Nawar Arony and Kezia Devathasan and Neil A. Ernst and Alexandra Branzan Albu and Daniela E. Damian},
  title = {A Data-Driven Approach for Finding Requirements Relevant Feedback from {TikTok} and {YouTube}},
  booktitle    = {Proceedings of the 31st IEEE International Requirements Engineering Conference (RE)},
  month = {September},
  pages = {111--122},
  year = {2023},
  address = {Hannover, Germany},
  url = {https://doi.org/10.1109/RE57278.2023.00020}
}

@inproceedings{Nayebi-RE23,
  author = {Maleknaz Nayebi and Konstantin Kuznetsov and Andreas Zeller and G{\"{u}}nther Ruhe},
  title = {User Driven Functionality Deletion for Mobile Apps},
  booktitle    = {Proceedings of the 31st IEEE International Requirements Engineering Conference (RE)},
  month = {September},
  pages = {6--16},
  year = {2023},
  address = {Hannover, Germany},
  url = {https://doi.org/10.1109/RE57278.2023.00011}
}

@inproceedings{Fantechi-RE23,
  author = {Alessandro Fantechi and Stefania Gnesi and Lucia C. Passaro and Laura Semini},
  title = {Inconsistency Detection in Natural Language Requirements using {ChatGPT}: A Preliminary Evaluation},
  booktitle    = {Proceedings of the 31st IEEE International Requirements Engineering Conference (RE)},
  month = {September},
  pages = {335--340},
  year = {2023},
  address = {Hannover, Germany},
  url = {https://doi.org/10.1109/RE57278.2023.00045}
}

@inproceedings{Ruhe-SEKE02,
	title        = {Quantitative {WinWin}: A new method for decision support in requirements negotiation},
	author       = {G{\"{u}}nther Ruhe and Armin Eberlein and Dietmar Pfahl},
	year         = {2002},
	month        = {July},
	booktitle    = {Proceedings of the 14th International Conference on Software Engineering and Knowledge Engineering},
	address      = {Ischia, Italy},
	series       = {SEKE'02},
	pages        = {159--166},
	doi          = {https://doi.org/10.1145/568760.568789}
}

@inproceedings{Palomba-ICSME15,
  author = {Fabio Palomba and Mario Linares V{\'{a}}squez and Gabriele Bavota and Rocco Oliveto, Massimiliano {Di Penta} and Denys Poshyvanyk and Andrea {De Lucia}},
  title = {User reviews matter! {T}racking crowdsourced reviews to support evolution of successful apps},
  booktitle    = {Proceedings of the 31st IEEE International Conference on Software Maintenance and Evolution (ICSME)},
  month = {September-October},
  pages = {291--300},
  year = {2015},
  address = {Bremen, Germany},
  url = {https://doi.org/10.1109/ICSM.2015.7332475}
}

@inproceedings{Palomba-ICSE17,
  author = {Fabio Palomba and Pasquale Salza and Adelina Ciurumelea and Sebastiano Panichella and Harald C. Gall and Filomena Ferrucci and Andrea {De Lucia}},
  title = {Recommending and localizing change requests for mobile apps based on user reviews},
  booktitle = {Proceedings of the 39th IEEE/ACM International Conference on Software Engineering (ICSE)},
  month = {May},
  pages = {106--117},
  year = {2017},
  address = {Buenos Aires, Argentina},
  url = {https://doi.org/10.1109/ICSE.2017.18},
}

@inproceedings{Ameller-PROFES16,
	title        = {A Survey on Software Release Planning Models},
	author       = {David Ameller and Carles Farr{\'{e}} and Xavier Franch and Guillem Rufi{\'{a}}n},
	year         = {2016},
	month        = {November},
	booktitle    = {Proceedings of the 17th International Conference on Product-Focused Software Process Improvement (PROFES)},
	address      = {Trondheim, Norway},
	pages        = {48--65},
	url          = {https://doi.org/10.1007/978-3-319-49094-6_4}
}

@inproceedings{Karim-SSBSE14,
	title        = {Bi-objective Genetic Search for Release Planning in Support of Themes},
	author       = {Muhammad Rezaul Karim and G{\"{u}}nther Ruhe},
	year         = {2014},
	month        = {August},
	booktitle    = {Proceedings of the 6th International Symposium on Search-Based Software Engineering (SSBSE)},
	address      = {Fortaleza, Brazil},
	pages        = {123--137},
	url          = {https://doi.org/10.1007/978-3-319-09940-8_9}
}

@inproceedings{Zhang-GECCO07,
	title        = {The multi-objective next release problem},
	author       = {Yuanyuan Zhang and Mark Harman and S. Afshin Mansouri},
	year         = {2007},
	month        = {July},
	booktitle    = {Proceedings of the 9th Annual Conference on Genetic and Evolutionary Computation (GECCO)},
	address      = {London, UK},
	pages        = {1129--1137},
	url          = {https://doi.org/10.1145/1276958.1277179}
}

@inproceedings{Berry-RE17,
  author = {Daniel M. Berry and Jane Cleland-Huang and Alessio Ferrari and Walid Maalej and John Mylopoulos and Didar Zowghi},
  title = {Panel: Context-Dependent Evaluation of Tools for {NL RE} Tasks: Recall vs. Precision, and Beyond},
  booktitle    = {Proceedings of the 25th IEEE International Requirements Engineering Conference (RE)},
  month = {September},
  pages = {570--573},
  year = {2017},
  address = {Lisbon, Portugal},
  url = {https://doi.org/10.1109/RE.2017.64}
}

@article{Robinson-CSUR03,
  author = {William N. Robinson and Suzanne D. Pawlowski and Vecheslav Volkov},
  title = {Requirements interaction management},
  journal = {ACM Computing Surveys},
  volume = {35},
  number = {1},
  month = {March},
  year = {2003},
  pages = {132--190},
  url = {https://doi.org/10.1145/857076.857079},
}

@article{Hindle2015,
  author    = {Abram Hindle and Christian Bird and Thomas Zimmermann and Nachiappan Nagappan},
  title     = {Do topics make sense to managers and developers?},
  journal   = {Empirical Software Engineering},
  volume    = {20},
  number    = {2},
  pages     = {479--515},
  year      = {2015},
  publisher = {Springer},
  doi       = {10.1007/s10664-014-9312-1},
  url       = {https://doi.org/10.1007/s10664-014-9312-1}
}

@inproceedings{Tiarks2014,
  author    = {Rebecca Tiarks and Walid Maalej},
  title     = {How Does a Typical Tutorial for Mobile Development Look Like?},
  booktitle = {Proceedings of the 11th Working Conference on Mining Software Repositories (MSR 2014)},
  pages     = {272--281},
  year      = {2014},
  publisher = {Association for Computing Machinery},
  address   = {New York, NY, USA},
  doi       = {10.1145/2597073.2597106},
  url       = {https://doi.org/10.1145/2597073.2597106},
  isbn      = {9781450328630},
  location  = {Hyderabad, India},
}

@article{Wang_TSE_22,
author = {Wang, Wentao and Dumont, Faryn and Niu, Nan and Horton, Glen},
title = {Detecting Software Security Vulnerabilities Via Requirements Dependency Analysis},
year = {2022},
issue_date = {May 2022},
publisher = {IEEE Press},
volume = {48},
number = {5},
issn = {0098-5589},
url = {https://doi.org/10.1109/TSE.2020.3030745},
doi = {10.1109/TSE.2020.3030745},
journal = {IEEE Trans. Softw. Eng.},
month = may,
pages = {1665–1675},
numpages = {11}
}

@article{Vargha_Delaney,
 ISSN = {10769986, 19351054},
 URL = {http://www.jstor.org/stable/1165329},
 author = {András Vargha and Harold D. Delaney},
 journal = {Journal of Educational and Behavioral Statistics},
 number = {2},
 pages = {101--132},
 publisher = {[American Educational Research Association, Sage Publications, Inc., American Statistical Association]},
 title = {A Critique and Improvement of the "CL" Common Language Effect Size Statistics of McGraw and Wong},
 urldate = {2026-07-19},
 volume = {25},
 year = {2000}
}

\end{document}